\documentclass{iopjournal-Copie}

\usepackage[backend=biber,style=numeric,sorting=none]{biblatex}
\usepackage{graphicx}
\usepackage{subcaption}
\usepackage[title]{appendix}
\usepackage{amsmath}

\begin{document}


\title{First Experimental Characterization of Plasma Parameters and Carbon Decontamination Rates in a Microwave Resonator Used in Particle Accelerators}

\author{Camille CHENEY$^1$, Gabriel ABI-ABBOUD$^{1}$, Stéphane BÉCHU$^{2}$, Alexandre BÈS$^{2}$, Laurent BONNY$^{2}$, Thibaut Gerardin$^1$, Bruno MERCIER$^1$, Eric MISTRETTA$^1$, Jonathan YEMANE$^1$ and David LONGUEVERGNE$^1$ }

\affil{$^1$Laboratoire de Physique des 2 Infinies Irène Joliot-Curie -- IJCLab, CNRS, Université Paris-Saclay, Orsay, France}

\affil{$^2$Laboratoire de Physique Subatomique et Cosmologie (LPSC), Univ. Grenoble Alpes, CNRS, Grenoble INP* (*Institute of Engineering Univ. Grenoble Alpes), LPSC-IN2P3, Grenoble, France}

\email{cheney@ijclab.in2p3.fr}


\begin{abstract}
In-situ plasma processing of superconducting radio frequency (SRF) cavities is a performance recovery technique used to mitigate the field emission limiting phenomenon. It has been proved very effective at major particle accelerator facilities such as SNS, CEBAF, FRIB, FNAL and C-ADS. 
This technique is based on the ignition of a noble-gas/oxygen plasma inside the cavity over several hours to remove hydrocarbon-based contamination, responsible for the parasitic field emission degradation observed after several years of operation. Despite a large experimental R\&D effort from the community, plasma parameters and cleaning rates under various experimental conditions have never been directly evaluated. In this study, plasma parameters were measured using a Langmuir probe and cleaning rates thanks to a quartz crystal microbalance (QCM) coated with an amorphous carbon film to simulate a carbon-based contamination. In this article, the main results from a large parameter space are discussed along with guidelines for improving the plasma processing effectiveness in SRF cavities. The encountered technical challenges are also discussed, as the SRF cavity is by design not intended to be a plasma reactor. 


\end{abstract}

\section{Introduction}
Radio frequency cavities are at the heart of a particle accelerator. They consist in a metallic resonator in which electric fields are amplified to several tens of MV/m. Acceleration of synchronized charged particles is achieved thanks to Lorentz forces. 

 SRF cavities are used in modern particle accelerators as the superconducting technology allows improved performances -- contrary to the normal conducting technology -- in some applications such as high power (MW range), high beam current (mA range) and continuous wave operation (CW). To achieve the superconducting state, niobium-made cavities must be cooled down using liquid helium and must be thermally isolated to minimize the heat load from the environment (300 K) to the cold mass (usually 4.2 K or 2 K). To fulfill these requirements, they are encapsulated in very complex cryostats called cryomodules.  

As continuously exposed to beam losses and warm parts (safety valves, beam pipes and diagnostics), SRF cavities experience a gradual drop in performance caused by the emergence or strengthening of the parasitic field emission phenomenon. This phenomenon arises from surface contamination that facilitates the tunneling and acceleration of electrons subject to electromagnetic fields, resulting in the production of ionizing X-rays. This not only raises safety concerns but also increases the thermal load within the liquid helium bath. Addressing this issue typically necessitates disassembling the accelerator cryomodule to reprocess the accelerating cavities. This complex operation requires the disassembly of the cryomodule, followed by various wet cleaning processes. Finally, cavities are assembled in a clean room before being reinstalled in the cryomodule. This operation is time consuming, labor intensive and expensive. As a last chance before disassembly, the SRF community has developed a specific in-situ recovery process, relying on plasma-surface interactions.

This new approach involving a plasma discharge has emerged in the 2010's at SNS \cite{Doleans2013PlasmaPR, Tyagi:2014bwa, 10.1063/1.4972838}. This technique consists in generating and sustaining over several hours a reactive plasma by injecting a gas mixture of a noble gas (He, Ne or Ar) with a few percentage of oxygen, and by igniting and maintaining the plasma discharge thanks to the RF system already installed in the cryomodule. Despite its confusing name, the plasma generated in the cavity is a wave driven discharge and not an RF discharge, in the sense of a capacitively coupled discharge. This "in-situ" treatment has proven to be effective in mitigating the field emission phenomenon \cite{martinello2022plasmaprocessinginsitufield, TYAGI201629} but without fully recovering initial performances.

Indeed, this technique shows significant limitations for several reasons. First, the RF system and reactor configurations are optimized for beam acceleration, but not for plasma generation. Secondly, some types of contamination (metallic flakes, dust) responsible of field emission are not as removed as carbon-based contaminants are in the state of the art.

In an effort to optimize and extend the capabilities of the plasma cleaning process in these resonator-type reactors, a specific setup was developed at IJCLab. It comprises a SPIRAL2 $\beta=0.12$ prototype cavity \cite{OLRY2006197} connected to a vacuum system allowing gas injection and flow through the cavity, and an RF system to provide electromagnetic power to the cavity. For the first time in this type of application, various plasma diagnostics have been installed in the setup, combining a Langmuir probe for plasma parameters measurements and a quartz crystal microbalance (QCM) coated with amorphous carbon to measure cleaning rates (see Fig. \ref{fig_setup}).

\begin{figure}[htb!]
 \centering
        \includegraphics[width=0.8\textwidth]{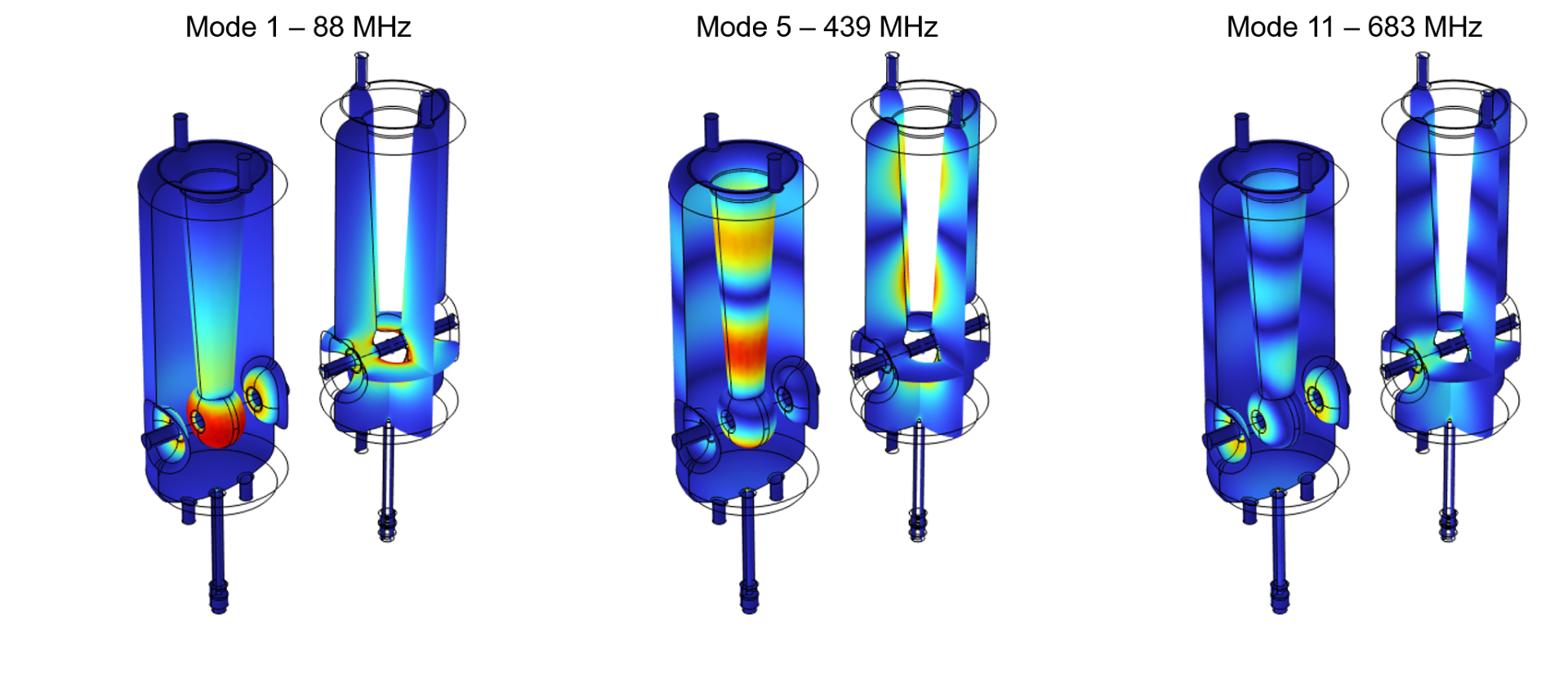}
 \caption{Electric field distribution for various resonant modes called higher-order modes (HOMs). Every mode has a unique E-field distribution, allowing the plasma to ignite in an area of interest by choosing the right HOM. For example, the 5th HOM has a high electric field area in the middle-height of the cavity, leading the plasma to ignite there in a toroidal shape, as the E-field. RF simulations were performed with COMSOL Multiphysics \cite{comsol}}
\label{fig1}
\end{figure}

Plasma generation within the cavity was thoroughly investigated prior to the study presented here. Plasma can be ignited and maintained in several locations in the cavity by driving the plasma with different resonant modes: the fundamental mode, first mode at 88 MHz and all the other at higher frequencies, called higher-order modes (HOMs), as shown in Fig. \ref{fig1}. 
Indeed, since each HOM exhibits a distinct electric field distribution, the plasma discharge can be spatially confined to a specific region of interest, effectively following the field’s spatial profile.
During this preliminary study, a mode was identified to be particularly interesting as it ignites and maintains the plasma in a region where many vacuum ports are in line of sight of the plasma. This combination of advantageous plasma location and cavity geometry made this study possible. 

\section{Method}
In this section, the experimental apparatus is presented with the testing setup elements along with diagnostic tools and their associated limitations.

\subsection{Vacuum System}
\begin{figure}[htb!]
 \centering
        \includegraphics[width=0.5\textwidth]{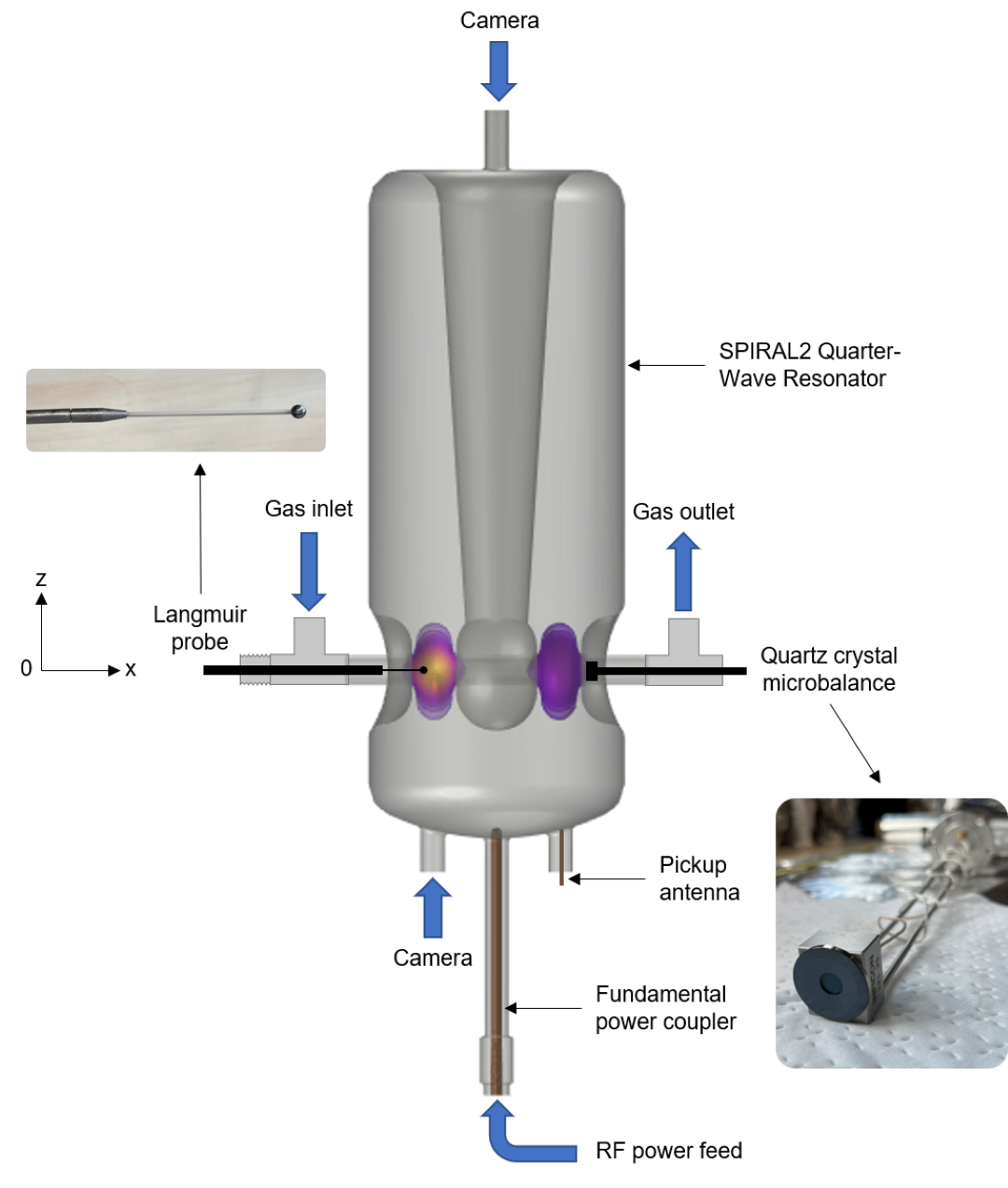}
 \caption{Schematic of the experimental setup. The Langmuir probe can be moved on the $x$ axis, while the quartz crystal microbalance is locked in a fixed position for all experiments. Cameras are installed at the top and bottom parts of the cavity to allow a qualitative diagnostic of plasma distribution. In this case, the plasma is ignited in both accelerating gaps, thus benefiting from the numerous vacuum ports providing access to the plasma for diagnostics.}
\label{fig_setup}
\end{figure}

For this study, a standard plasma processing vacuum system was used. It consists, on one side of the cavity, in a gas injection system made of two mass flow controllers (MFC). One MFC is dedicated to a dilution gas (He, Ar, N$_2$) ranging from 2.4 sccm to 500 sccm. The second MFC ranging from 0.04 sccm to 2.2 sccm is dedicated to O$_2$. On the other side of the cavity, it consists in a pumping system composed of a 56 L/s turbo-molecular pump backed with a rotary vane roughing pump. In parallel, a residual gas analyzer (RGA) with differential pumping is implemented on the gas outlet side. It is used to monitor over time the species created or consumed when plasma is ignited. To monitor the vacuum level, a set of Pirani, Penning and capacitive gauges were used. The latter are of primary importance as they allow accurate pressure readings, independently of the type of gas, as several gas mixtures were investigated. Fig. \ref{fig_setup} shows a schematic diagram of the experimental setup.

\subsection{RF System}
The RF system consists in a signal generator [125 kHz -- 1024 MHz] connected to a broad band amplifier [80 MHz -- 1 GHz] with a maximum output power of 250 W. The RF amplifier can withstand full reflection. A bi-directional coupler is placed on the RF line to monitor the forward power ($P_{FWD}$) and the reverse power ($P_{REV}$). Finally, the RF line is connected to the cavity's fundamental power coupler (FPC) \cite{GOMEZMARTINEZ201737}, as used during standard accelerator operation to mimic real mismatched conditions inducing technical limitations as it will be described later. A pickup probe antenna is placed on the cavity to monitor the electric field magnitude by measuring the transmitted power ($P_t$) and to determine the various cavity resonance frequencies. RF power measurements are performed with power-meters. A vector network analyzer (VNA) was occasionally used to monitor the scattering parameters ($S_{11}$ and $S_{21}$) with and without plasma.

\subsection{Langmuir probe}

In this subsection, the theoretical framework of Langmuir probe diagnostics will be briefly outlined, along with the measurement hardware used in this study. The discussion will also address the challenges posed by plasma instabilities induced by probe insertion within the RF cavity plasma reactor.

\subsubsection{Langmuir probe overview}

Langmuir probes are a widely used diagnostic tool in plasma physics, enabling the measurement of key plasma parameters such as electron density ($n_e$), electron temperature ($T_e$), plasma potential ($V_p$), and the Electron Energy Distribution Function (EEDF).

In a plasma, electrons exhibit a distribution of velocities rather than a single kinetic energy, analogous to a gas. This distribution is characterized by the EEDF, $F_0(\varepsilon)$, which represents the probability density of electrons as a function of their energy, $\varepsilon$. Unlike the electron energy probability function (EEPF), the EEDF is normalized by the electron density, $n_e$, such that:

\begin{equation}
\int_0^{+\infty} \varepsilon^{1/2} F_0(\varepsilon) \, d\varepsilon = 1
\end{equation}

The so-called electron temperature, $T_e$, is actually an effective electron temperature, defined as 2/3 times the electron mean energy of the distribution, as the following equation:

\begin{equation}
T_e = \frac{2}{3} \int_0^{+\infty} \varepsilon^{3/2} F_0(\varepsilon) \, d\varepsilon
\label{equation2}
\end{equation}


Since experimental EEDFs often deviate from a Maxwellian distribution -- evident as a linear trend in a semilogarithmic plot of EEDF versus energy ($\varepsilon$) -- it is common to define multiple temperatures for distinct energy ranges.
For instance, in the case of a bi-Maxwellian EEDF, two characteristic temperatures can be identified: $T_{\text{e, low}}$ and $T_{\text{e, high}}$, corresponding to the low-energy electron population and the high-energy  population, respectively -- also called high-energy tail.
An illustrative example of the assignment of $T_{\text{e, low}}$ and $T_{\text{e, high}}$ for an Ar/O$_2$ plasma is provided in Fig.~\ref{fig13b}.

All electron densities reported in this study were derived from the EEPF, $f(\varepsilon)$, using the following formula:

\begin{equation}
n_e = \int_0^{+\infty} \varepsilon^{1/2} f(\varepsilon) \, d\varepsilon
\end{equation}

For this study, Langmuir probes were custom-designed and fabricated to meet specific requirements, such as relatively low plasma densities. A variety of cylindrical and spherical probes with different diameters were tested, with spherical probes proving most suitable for the unique conditions of our plasma reactor. The measurements presented here were conducted using two spherical probes with diameters of 6 mm and 10 mm.


All I–V characteristics were processed using a custom Python implementation of the Druyvesteyn method, which extracts plasma parameters from the second derivative of the I–V curve \cite{druyvesteyn1930niedervoltbogen, K_Popov_2006, BechuEtudeSonde}. This approach is particularly robust because, under the usual probe assumptions, it provides reliable plasma parameters for arbitrary EEDF shapes and for any convex probe surface \cite{chen2003langmuir}. In addition, spherical probes help mitigate the effects of velocity‑space anisotropy, since their symmetry ensures an orientation‑independent effective collection area and thus reduces directional bias in the inferred EEDF \cite{10.1063/1.1149862}.

The probe was aligned along the cavity beam axis and could be inserted at different depths, facilitating spatially resolved measurements. This was very useful to crosscheck measurements with numerical simulations and was also crucial to determine the optimal probe position. This topic is discussed in detail in appendix \ref{appendixA}.

Three distinct probe electronics systems were employed to perform the measurements. Preliminary data were acquired using the "Quë-Do" electronics developed in-house at LPSC \cite{Plaquette_Que_Do}. Subsequently, an Impedans ALP System, provided by the Laboratoire de Physique des Plasmas (LPP), was utilized. Finally, a custom-built system based on a Keithley 2400 sourcemeter, controlled via a Python script, was also employed. All three systems yielded consistent plasma parameters for a given plasma condition.

\subsubsection{Plasma instabilities caused by the Langmuir probe} \label{subsect4.1.1}
\begin{figure}[htb!]%
    \centering
    \subfloat[\centering Electric field distribution of the HOM at 683 MHz]{{\includegraphics[width=0.5\textwidth]{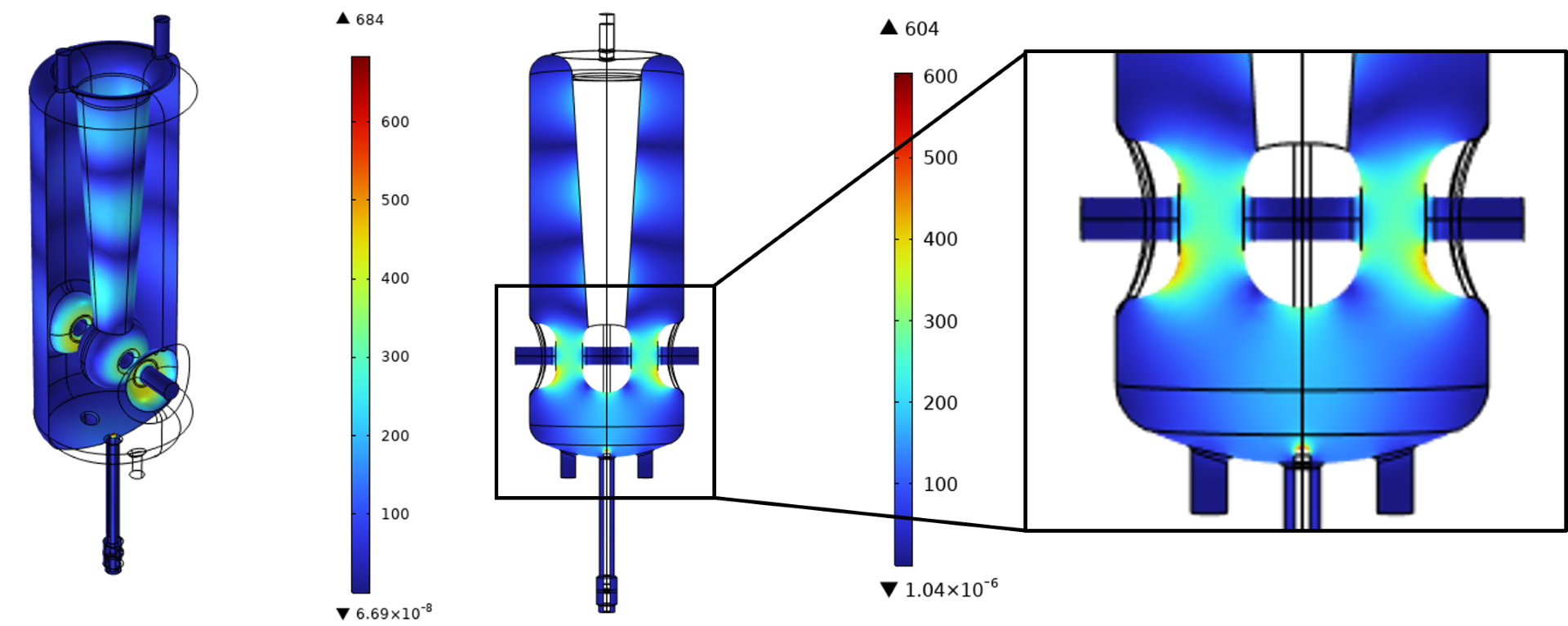} }}%
    \qquad
    \subfloat[\centering Pictures from the plasma light for various gas mixtures]{{\includegraphics[width=0.5\textwidth]{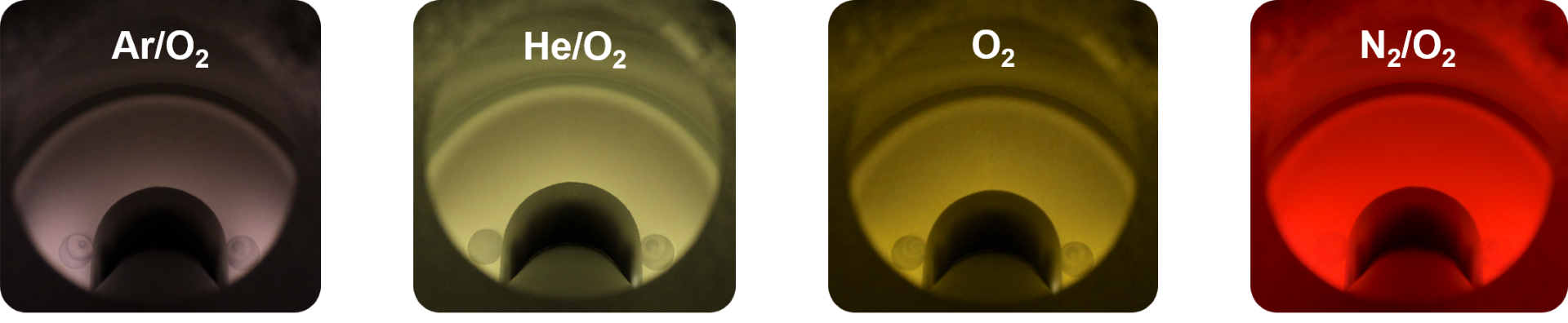} }}%
    \caption{Electric field distribution and plasma light associated to the 11th HOM at 683 MHz}%
    \label{fig3}
\end{figure}

Due to the fact that our plasma reactor is a resonator, inserting a Langmuir probe in the cavity volume will inevitably perturb the electromagnetic field. Furthermore, the probe acts as an antenna, picking RF power, and enhancing the electric field around the probe's tip. To overcome this issue, the plasma should be located close to a vacuum port, minimizing the probe length inserted inside the cavity volume. The 683 MHz HOM (see Fig. \ref{fig3}) fulfills these requirements, igniting the plasma in the accelerating gaps region. This is also very convenient for the cleaning rate study with the QCM as both the Langmuir probe and QCM are facing the plasma from the same location, but on the opposite and symmetrical side of the cavity.

Independently from the gas mixture, the plasma will always ignite in both accelerating gaps, see Fig. \ref{fig3} . However for Ar, Ar/O$_2$, N$_2$ and N$_2$/O$_2$ mixtures the plasma will extinguish in one accelerating gap -- while remaining ignited in the other gap -- during the frequency tuning process. This phenomenon is still not very well understood, but plasma oscillations are detected when approaching the two gaps to one gap transition, with the frequency of these oscillations decreasing when approaching the transition. The gap in wich the plasma will remain ignited is random. This loss of plasma distribution symmetry is not observed for pure O$_2$ and He/O$_2$ mixtures.

Furthermore, for all gas mixtures, attempts to perform probe measurements with the plasma ignited in both accelerating gaps revealed a persistent instability.
Specifically, when the probe bias approached the plasma potential during voltage scans, a sudden jump in the I-V characteristic was observed, rendering the analysis of probe curves unfeasible due to the inability to determine the plasma potential ($V_p$).
This instability was found to be strongly dependent on probe location: the deeper the probe insertion, the more pronounced the effect.
In extreme cases, pushing the probe too far into the plasma caused a secondary plasma to ignite around it. This made the measurements useless, as they ended up describing the secondary plasma instead of the primary discharge.

To mitigate this issue, positioning the probe flush with -- or slightly retracted from -- the accelerating gap proved effective.
However, this adjustment meant the probe was no longer situated at the core of the plasma bulk, where electron density and temperature are expected to peak.
As a result, the measured electron density in this configuration was approximately two times lower than the peak density.
Further details on the axial distribution of plasma parameters are provided in Appendix \ref{appendixA}, which includes experimental results, numerical simulations using a plasma fluid model, and an electron kinetic interpretation.

To conclude, many difficulties were faced when trying to perform reliable and meaningful probe measurements. Several issues arose, once again, due to the specific characteristics of the SRF cavity plasma reactor. Solving those difficulties represented itself a thorough preliminary study.

\subsection{Quartz Crystal Microbalance (QCM)}
Quartz crystal microbalances (QCM) are typically used in thin film deposition systems such as physical, chemical vapor deposition (PVD, CVD) or atomic layer deposition (ALD) to monitor the growth of the thin film. For the purpose of this study, the QCM was used in a reversed way. Indeed, the QCM (INFICON Front Load Single crystal sensor) was coated prior to plasma exposure. The coating is amorphous carbon (a-C) deposited on the QCM head using a carbon thread evaporation system (safematic CCU-010 HV\footnote{with the CT-010 carbon thread evaporation module}) -- this is done outside of the cavity. After coating, the QCM is installed on the cavity beam axis -- like the Langmuir probe -- using the second vacuum port available. The deposited a-C thickness was typically between 50 nm and 200 nm.





It is important to note that the QCM is positioned to directly face the plasma, exposing its sensor to fluxes of neutral species, ions, and electrons. Thus, thermal stability was a key consideration in the experimental setup. No water cooling was required for the QCM, as the quartz sensor exhibited negligible heating during plasma exposure. Temperature measurements, recorded via a thermocouple inserted in the cooling pipes, confirmed a temperature increase of less than 2°C over more than one hour of continuous plasma discharge. This minimal thermal change is particularly significant given the sensitivity of AT-cut quartz crystals, which can drift by up to 10 Hz/°C \cite{inficon}, which is equivalent to 0.54 nm/°C for an amorphous carbon coating. Since each removal rate measurement involved the removal of several nanometers of material, any error introduced by temperature fluctuations is negligible.

Furthermore, the carbon removal rate remained constant under fixed discharge conditions. If sensor heating had occurred, it would have manifested as a non-linear removal rate over time, given the quartz's specific temperature sensitivity. The absence of such behavior confirms that thermal effects did not influence the measurements. Additionally, when the plasma discharge was stopped, the QCM frequency remained stable over several tens of minutes, indicating no post-discharge cooling of the sensor. This further validates the thermal stability of the system.


\section{Control Parameters and Limitations}
The plasma could be controlled by the following parameters:
\begin{itemize}
\item Resonant mode (HOM),
\item RF power,
\item DC bias applied on the power coupler,
\item Frequency tuning after plasma ignition,
\item Dilution gas (He, Ar or N$_2$),
\item O$_2$ concentration,
\item Gas pressure,
\end{itemize}

In this section, control parameters and their associated technical or physical limitations will be discussed to provide an overview of what can be achieved with this very uncommon plasma reactor.

\subsection{Resonant mode (HOM) driving the plasma} \label{section3.1}
In the context of RF cavity systems, coupling refers to the efficiency with which power is transferred from the applicator -- in this case, the fundamental power coupler (FPC) -- to the resonant cavity. It is a critical parameter that determines the fraction of incident power effectively utilized for plasma generation, as opposed to being reflected back to the power source. The coupling between the FPC and the cavity is imposed by the design to ensure optimal coupling during beam operation at cryogenic temperature and constitutes a significant operational constraint at room temperature.

Since plasma processing is performed at room temperature -- a necessity to prevent gas condensation on the cavity’s inner surfaces -- the coupling remains particularly weak. Consequently, only a small fraction of the incident power is transmitted to the cavity, with the majority reflected back to the amplifier, resulting in a highly inefficient process. This coupling is quantitatively described by the ratio $Q_0/Q_{\text{ext}}=\beta_{\text{ext}}$, where $Q_0$ represents the intrinsic quality factor of the cavity and $Q_{\text{ext}}$ the power coupler quality factor. 

For a given mode, $\beta_{\text{ext}}$ can be $<1$ or $>1$. In order to achieve optimum power transmission to the cavity, the closer the value of $\beta_{\text{ext}}$ to unity, the better.

\begin{figure}[htb!]
 \centering
        \includegraphics[width=0.4\textwidth]{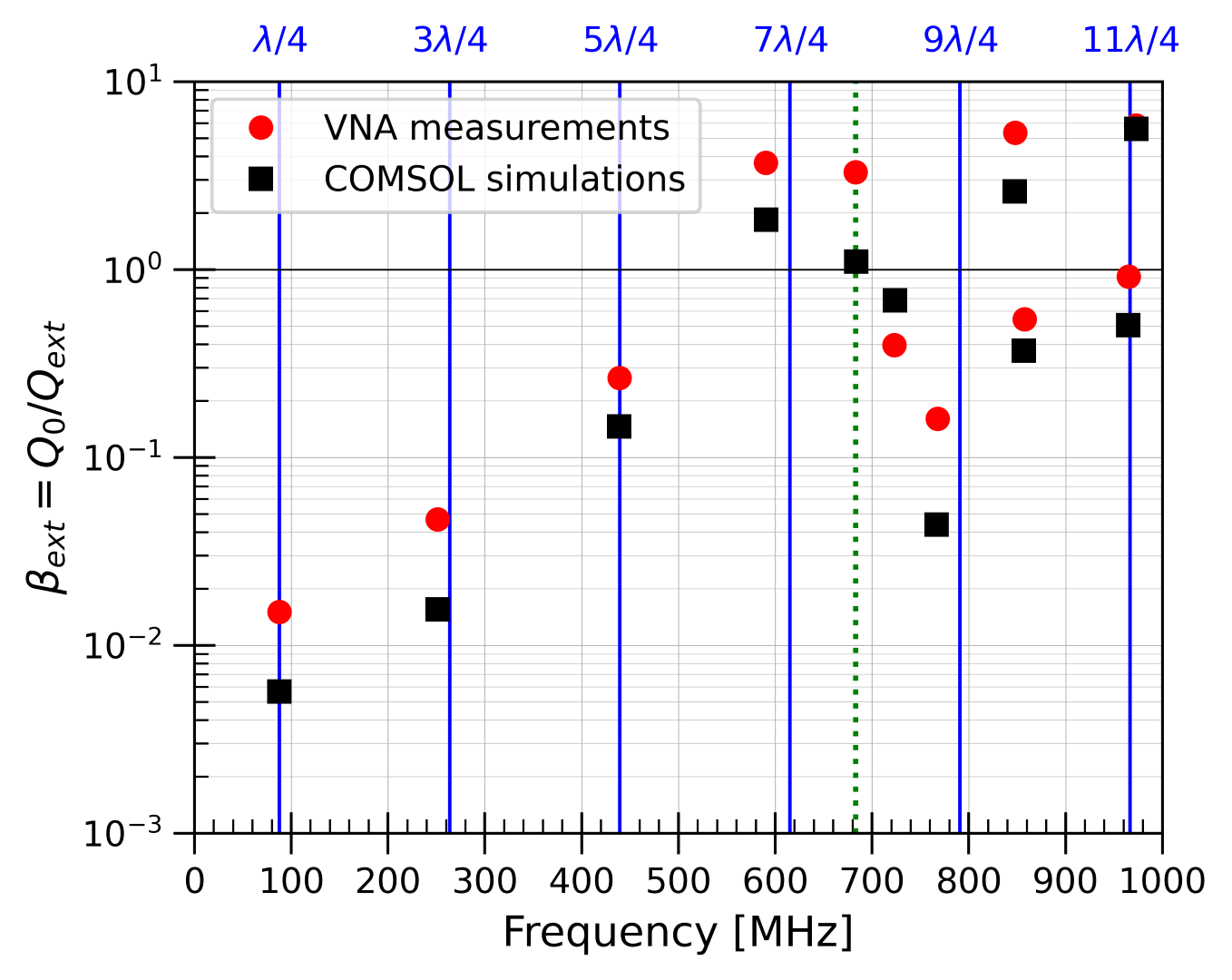}
 \caption{Coupling factor $\beta_{\text{ext}}$ as a function of several HOM frequencies. High-frequency HOMs are significantly better coupled than the fundamental mode at 88 MHz. The 683 MHz HOM is highlighted with the green dotted line, showing a 21 time increase in coupling compared to the fundamental mode.}
\label{fig1.1}
\end{figure}

For the SPIRAL2 cavity under investigation, the coupling factor $\beta_{\text{ext}} \approx 1.5\times10^{-2}$ for the fundamental mode at 88 MHz at room temperature. Thus, only a small fraction of the incident power is actually transmitted to the cavity. To mitigate this weak coupling limitation, the use of higher-frequency resonant modes -- higher-order modes (HOMs) -- is particularly advantageous. These modes significantly improve coupling, as illustrated in Fig. \ref{fig1.1}, while also enabling the transmission of higher RF power before reaching the coupler breakdown threshold. This effectively widens the usable RF power window (see Section~\ref{sec3.2} and Fig.~\ref{fig2}). Among these HOMs, mode 11 at 683 MHz stands out as the optimal choice: although it requires a higher ignition power compared to mode 1 at a given pressure, it offers significantly enhanced coupling -- transferring up to 21 times more power -- while maintaining a plasma spatial distribution similar to that of the fundamental mode.

However, the use of HOMs results in plasma ignition at higher power levels at a given pressure, despite being better coupled. Additionally, as the HOM frequency increases, the electric field distribution becomes more complex (see Fig. \ref{fig1}), making the control of plasma spatial distribution more challenging.

Furthermore, as can be seen in Fig. \ref{fig1.1}, there are more modes above 600 MHz and they are closer to each other in the frequency range. This has some consequences on the frequency tuning process with the ``mode transition'' phenomenon. It is further developed in section \ref{section3.4}.

These constraints, which do not typically arise in standard plasma reactors designed for predefined operating conditions, are inherently imposed by the cavity architecture itself.

\subsection{RF power} \label{sec3.2}
As for every gas discharge, a sufficiently high electric field is needed to breakdown the gas, then leading to plasma ignition. Due to the high quality factor -- even at room temperature --  of the SPIRAL2 cavity, a plasma can be ignited at very low power, below 1 W. The ignition threshold naturally depends on the gas pressure and nature, following a Paschen-like law. 

\begin{figure}[htb!]
 \centering
        \includegraphics[width=0.95\textwidth]{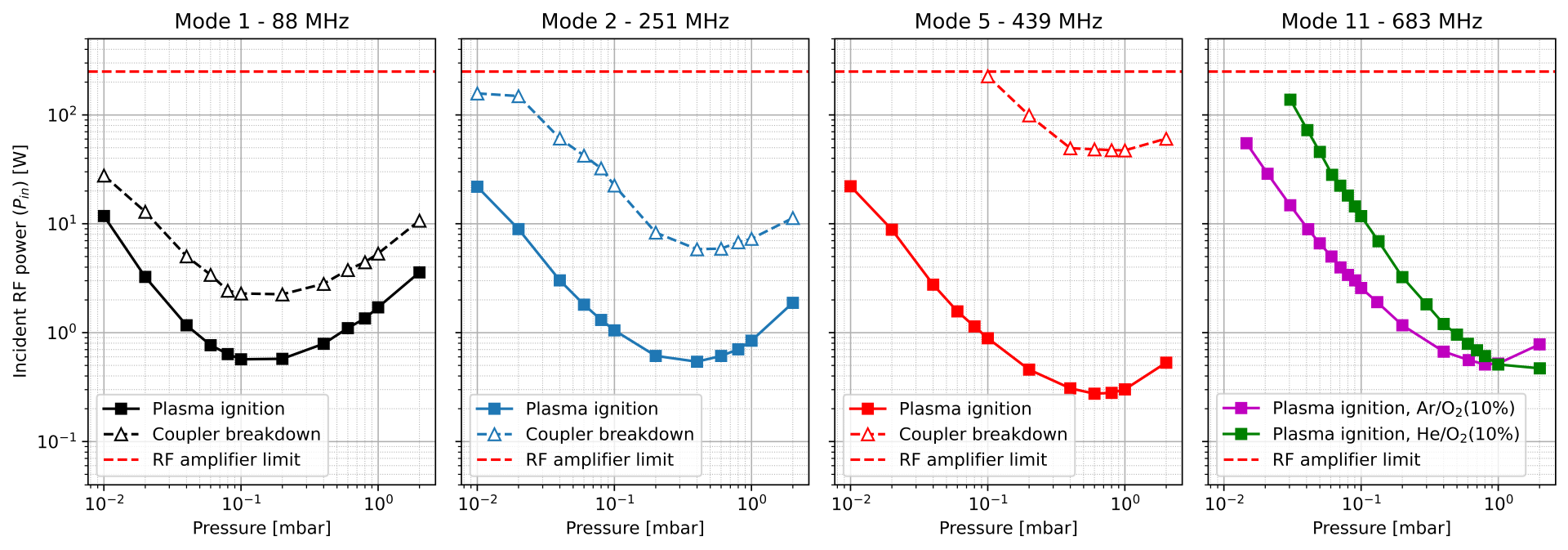}
 \caption{Experimental Paschen curves for the fundamental mode at 88~MHz, the second mode at 251~MHz, the fifth mode at 439~MHz, and the eleventh mode at 683~MHz, measured in an Ar/O$_2$(10\%) gas mixture. For comparison, data for a He/O$_2$(10\%) mixture are also shown for the 683~MHz mode. The solid line indicates the cavity plasma ignition threshold, while the dashed line represents the coupler breakdown threshold. The coupler breakdown threshold for the 683~MHz mode is not represented, as it could not be reached below $\approx 10^{-1}$ mbar.}
\label{fig2}
\end{figure}

Another peculiarity of the RF cavity plasma is ``coupler breakdown'', as usually called in the community. This phenomenon arises when a second RF power threshold is crossed. In this case, the plasma very quickly moves from the cavity volume to the FPC region \cite{cheney, cheneyLCWS}. Coupler breakdown is an unwanted phenomenon and must be avoided because copper from the FPC can be sputtered on top of the niobium or damage the ceramic window in the FPC, as reported here \cite{hartung:srf2023-thixa01, wu:srf2019-mop085}, leading to poor superconducting performance or damages. The experimental Paschen curves for four different modes with an Ar/O$_2$(10\%) mixture are represented in Fig. \ref{fig2}.

Coupler breakdown arises from the plasma-induced modification of the cavity's dielectric constant, which shifts the resonance frequency of the cavity ($f_0$) away from the driving frequency ($f_d$). As a result, the system experiences an increasing mismatch, causing the electric field to increase at the tip of the fundamental power coupler (FPC). When the power reaches a critical level, the localized electric field at the coupler tip becomes sufficiently large, leading a spatial redistribution of the plasma toward the coupler region. This change in the plasma distribution is rather similar to what Hagelaar and Hassouni reported in \cite{10.1063/1.1769607, Hassouni_2010}, although the gas heating is much less pronounced in our case.

To overcome this limitation, operating at a gas pressure approximately one order of magnitude below the Paschen minimum extends the usable power range between cavity plasma ignition and coupler breakdown. Additional strategies include frequency tuning to align the driving frequency with the plasma-perturbed resonance frequency (see section \ref{section3.4}), as well as applying a negative DC bias to repel electrons from the coupler region (see section \ref{sec:DC_bias}).

\subsection{DC bias of the powered antenna} \label{sec:DC_bias}
\begin{figure}[htb!]
 \centering
        \includegraphics[width=0.32\textwidth]{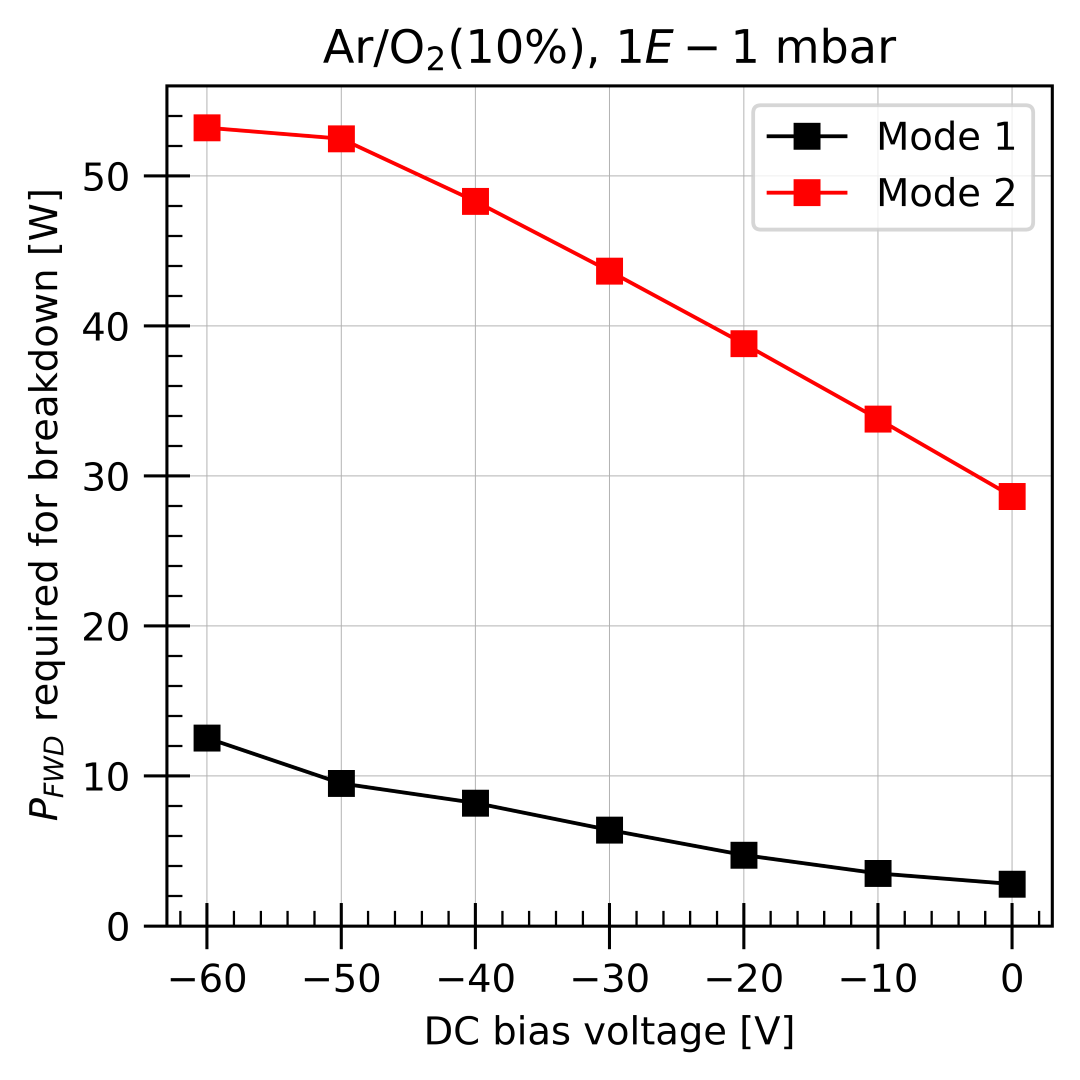}
 \caption{RF power required to induce coupler breakdown as a function of the DC bias applied to the fundamental power coupler (FPC) for an Ar/O$_2$(10\%) plasma. The data are presented for mode 1 at 88~MHz and mode 2 at 251~MHz. At a DC bias of $-60$~V, the breakdown threshold power increases by a factor of 4.5 for mode 1 and 1.9 for mode 2, highlighting the effectiveness of negative DC bias in delaying coupler breakdown.}
\label{fig2.2}
\end{figure}

Another leverage for delaying the coupler breakdown threshold consists in applying a DC bias to the powered antenna (FPC), using a bias-T. The bias-T allows to add a DC component to the RF signal. When a negative DC bias is applied to the FPC, electrons are repelled from the tip of the antenna, delaying the coupler breakdown apparition, see Fig. \ref{fig2.2} (adapted from \cite{cheney}). 
 
Inversely, when a positive bias is applied, electrons are attracted toward the antenna and the coupler breakdown occurs at even lower RF power. Indeed, it was observed that the linear trend exhibited in Fig. \ref{fig2.2} extends in the positive bias range.
 
A negative DC bias of a few tens of volts (-40V to -60V) seems to be sufficient to repel electrons and to have a comfortable safety margin \cite{cheney, hartung:hiat2025-tuc03}.

\subsection{Frequency tuning after cavity plasma ignition} \label{section3.4}
RF power controls the electric field amplitude and thus the electron density -- as in conventional plasma reactors. As mentioned in the previous section, the maximum amount of power transferred to the cavity is limited by the coupler breakdown threshold. This inevitably limits the maximum electron density achievable, consequently limiting the plasma cleaning effectiveness. 
To overcome this power limitation, the RF drive frequency can be tuned after plasma ignition. Indeed, as the dielectric constant decreases ($\varepsilon=1-\frac{\omega_p^2}{\omega^2}<1$) within the plasma volume, the resonant frequency is inevitably increased resulting in a significant mismatch between drive and actual resonant frequency. Tuning the drive frequency toward the resonant frequency increases the RF power transferred to the cavity (passband behaviour) leading to higher electron density. As dielectric constant is proportional to electron density ($\omega_p=\sqrt\frac{n_e~e^2}{\varepsilon_0~m_e}$), the resonant frequency is continuously increased during this process. By repeating the frequency increase process, the electron density can be significantly increased.

To summarize, electron density is controlled by tuning the frequency contrary to typical plasma reactors where $n_e$ is controlled by the RF power. In the next sections, the frequency increase will be referred as ``frequency tuning'', abbreviated with the symbol $\Delta f$ being the frequency difference between the vacuum resonance frequency of a given mode, and the drive frequency: $\Delta f = f_d - f_0$.

Nonetheless, the frequency tuning method has some limitations. Indeed, the maximum frequency shift achievable depends on:
\begin{itemize}
	\item Gas pressure,
	\item RF power,
	\item Resonant mode,
	\item Bandwidth of the resonant mode.
\end{itemize}

As a general rule, the maximum frequency shift is reached when: (1) the pressure is low, (2) the RF power is high, (3) the resonant mode is well coupled and (4) the bandwidth of the resonant mode is large. Due to these constraints, generating a high density plasma at low pressure ($1\times10^{-2}$ to $ 3\times10^{-2}$ mbar) is usually feasible. However, this cannot be achieved at higher pressure ($>1\times10^{-1}$ mbar) due to the combination of coupler breakdown limitations and the impossibility to increase the frequency, otherwise resulting in plasma extinction.

\begin{figure}[htb!]
 \centering
        \includegraphics[width=0.31\textwidth]{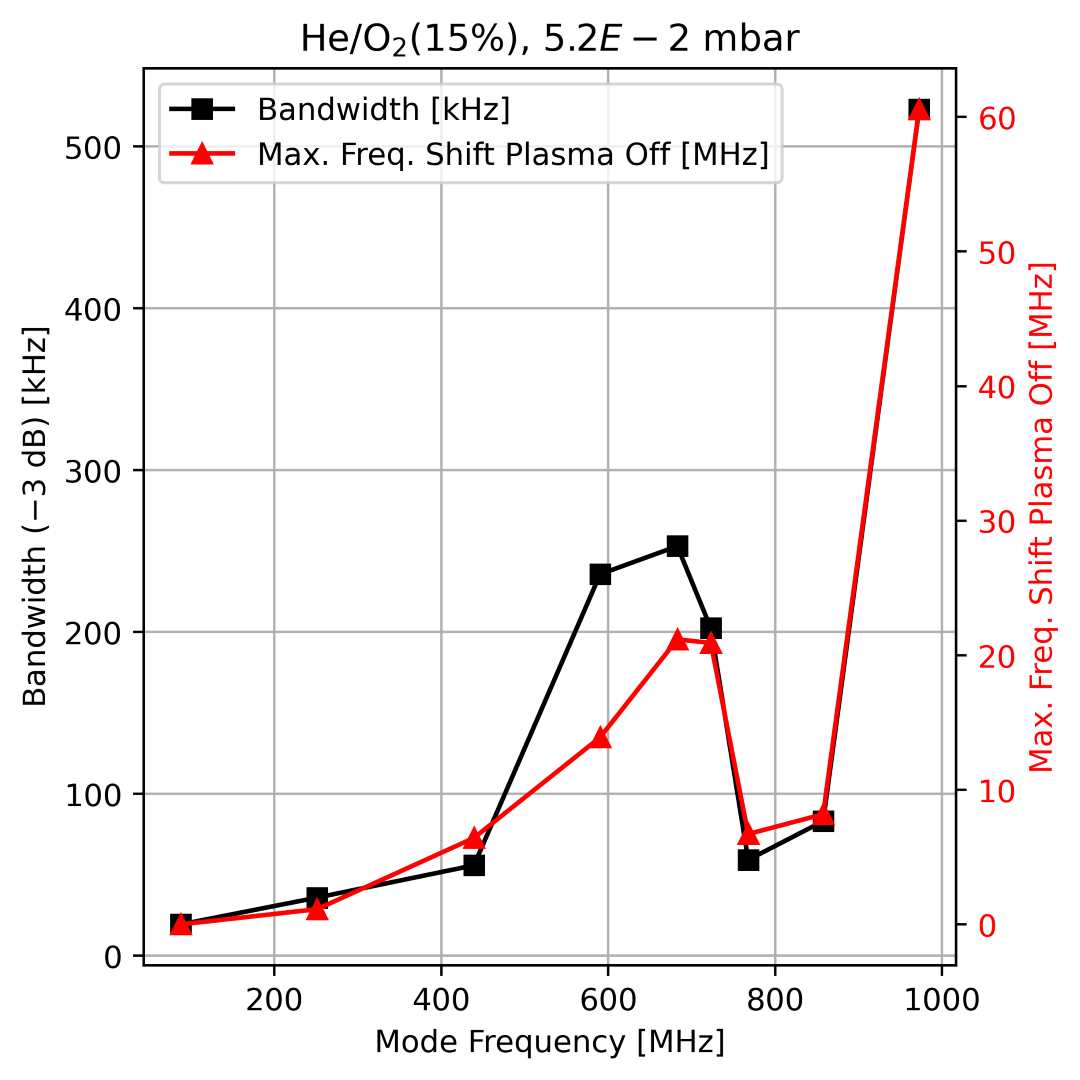}
 \caption{Bandwidth ($-3$ dB) of the various well-coupled HOMs and maximum frequency shift before plasma extinction. A good correlation between bandwidth and maximum frequency shift is observed. The larger the bandwidth, the higher the maximum frequency shift.}
\label{fig2.3}
\end{figure}

Fig. \ref{fig2.3} Shows the correlation between the bandwidth and the maximum frequency shift attainable for a set of relatively well-coupled HOMs. The measurement of the maximum frequency shift is taken with the following procedure, in order to perform measurement in similar conditions for every HOM: at a given gas pressure (here $5.2\times10{-2}$ mbar with He/O$_2$(15\%)), the RF power is slowly increased until plasma ignites. While keeping this ignition power constant, the frequency is increased until the plasma extinguish by itself. This extinction frequency is repeatable over multiple measurements. 

Finally, as introduced in section \ref{section3.1}, driving the plasma with higher-order modes above 600 MHz can lead to a phenomenon known as \textit{mode transition}.
This transition occurs when the frequency tuning process causes the initially driven HOM to cross and overlap with a neighboring HOM.
As a result, the electric field distribution changes, leading to a spatial redistribution of the plasma.
Depending on the application, this redistribution can be advantageous if the plasma relocates to a region of interest, or detrimental if it shifts to an undesirable area.
Additionally, the proximity of neighboring HOMs in the frequency space imposes a limit on the maximum achievable electron density for a given mode.

\subsection{Dilution gas and O$_2$ concentration}
In this study, various gas mixtures were investigated: pure He, Ar, N$_2$, O$_2$ and various oxygen based mixtures: He/O$_2$, Ar/O$_2$, N$_2$/O$_2$ while varying the O$_2$ concentration between 0\% and approximately 40\%. The maximum of 40\% is limited by the mass flow controllers. The oxygen concentration is set based on the oxygen over total gas flow ratio: 
\[
x_{O_2}[\%]= \frac{\Phi(O_2)}{ \Phi(O_2) + \Phi(He,~Ar,~N_2)} \times 100
\]

Where the gas flows $\Phi$ are given in sccm units. This method was found to be more accurate, reliable and repeatable than determining the oxygen concentration based on the oxygen partial pressure with the RGA. It is also much more convenient for the operator as the O$_2$ concentration can be predicted based on the gas flow set point. The impact of the gas mixture on the plasma parameters and cleaning rates will be discussed in this article. 


Lastly, the concentration of O$_2$ significantly influences the plasma ignition threshold. Especially, igniting a pure helium plasma demands high power, particularly at low pressures. However, introducing even a few percent of oxygen ($\approx5$\%) into helium appreciably reduces the plasma ignition threshold.

\subsection{Gas pressure}
Among all the control parameters, gas pressure is one of the most critical, controlling the overall plasma behavior. It dictates the power ignition and coupler breakdown thresholds, the maximum frequency shift $\Delta f$ achievable, hence the maximum electron density. It controls the plasma spatial distribution as well as electron, ion and neutral species mean free paths. 

As mentioned before, working at rather low pressure compared to the Paschen minimum not only decreases the coupler breakdown risk, but also increases the maximum electron density achievable -- through larger frequency shifts $\Delta f$. Nonetheless, if the pressure is too low, the ignition threshold increases even more and the plasma can be very difficult to ignite. In some cases, there is only coupler breakdown, without cavity plasma at too low pressure. This depends on the HOM being used.


Pressure also significantly influences the spatial distribution of the plasma. At lower pressures, the plasma tends to spread over a larger volume and allow a more spatially homogeneous distribution due to the increased mean free path of electrons. When the plasma occupies a larger volume, it interacts with a broader surface area, thereby enhancing local cleaning effectiveness -- even though plasma is known to exhibit long-range effects beyond its immediate location in the reactor, especially at lower pressures. Additionally, if the electric field pattern exhibits a dipole-like or quadrupole-like distribution, all poles are generally ignited at lower pressures. In contrast, at higher pressures, only a single pole tends to ignite. This homogenization effect is attributed to the longer electron mean free path at lower pressures.

Additionally, the pressure set point depends on two factors. First, the total gas flow and secondly the effective pumping speed at the cavity -- which depends on the turbo pump pumping speed and the conductance of the various vacuum equipment between the turbo pump and the cavity. Usually, this conductance is the limiting factor to achieve pressure below $10^{-2}$ mbar while still imposing a gas flow.

\section{Results}
In this section, the main findings will be discussed. This study focused on a large parameter space investigation aiming at understanding the mechanisms controlling the plasma in this unusual plasma reactor. The final goal is to provide an overview of the leverages to optimize the plasma processing for SRF cavities.

\subsection{Frequency tuning} \label{section4.3}
As discussed in section \ref{section3.4}, the frequency tuning is the way to compensate the frequency shift due to plasma ignition to have a more efficient power transmission to the plasma in the case of the SRF cavity plasma reactor, while significantly decreasing the risk of coupler breakdown. This technique is widely used in the community to enhance the plasma cleaning effectiveness by increasing the electron density. However, no direct measurements of plasma parameters have been performed in such a resonator before, so the electron density gain through this method was unknown. 

\begin{figure}[htb!]
\centering
     \begin{subfigure}[t]{0.49\textwidth}
         \centering
         \includegraphics[width=\textwidth]{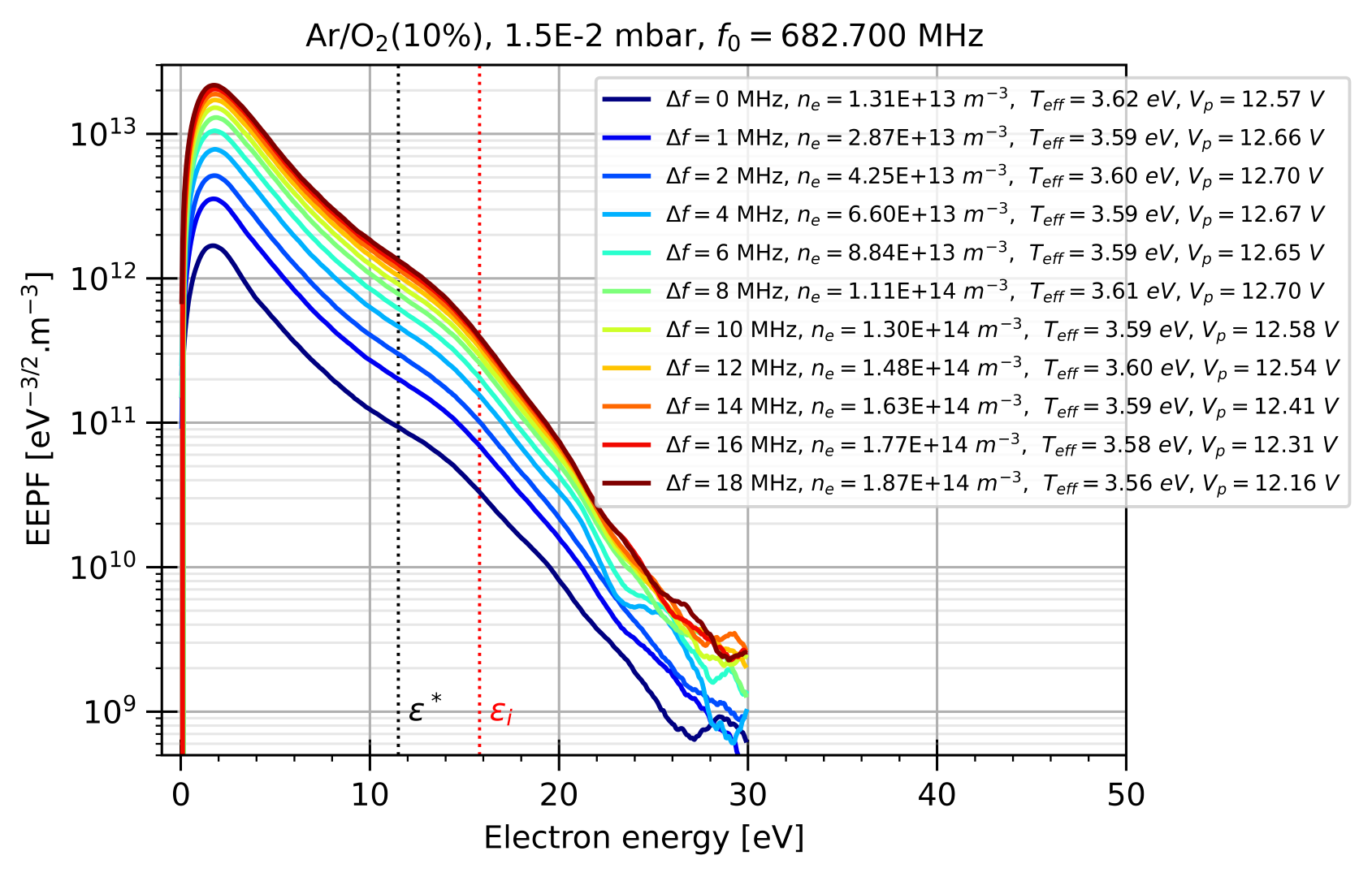}
         \caption{EEPF evolution with frequency shift}
         \label{fig7a}
     \end{subfigure}
      \hfill
     \begin{subfigure}[t]{0.49\textwidth}
         \centering
         \includegraphics[width=\textwidth]{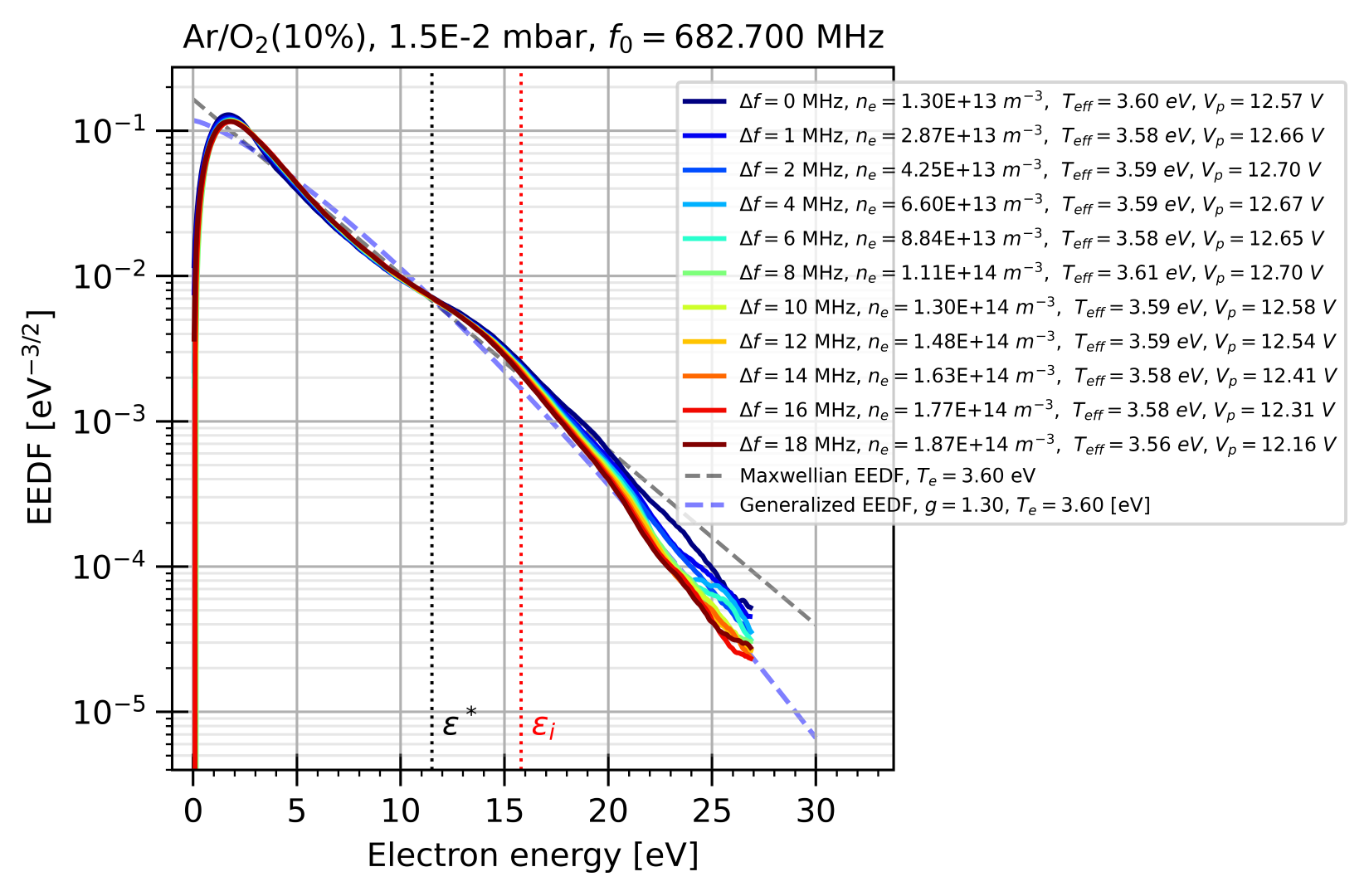}
         \caption{EEDF evolution with frequency shift}
         \label{fig7b}
     \end{subfigure}
     \hfill
     \begin{subfigure}[t]{0.45\textwidth}
         \centering
         \includegraphics[width=\textwidth]{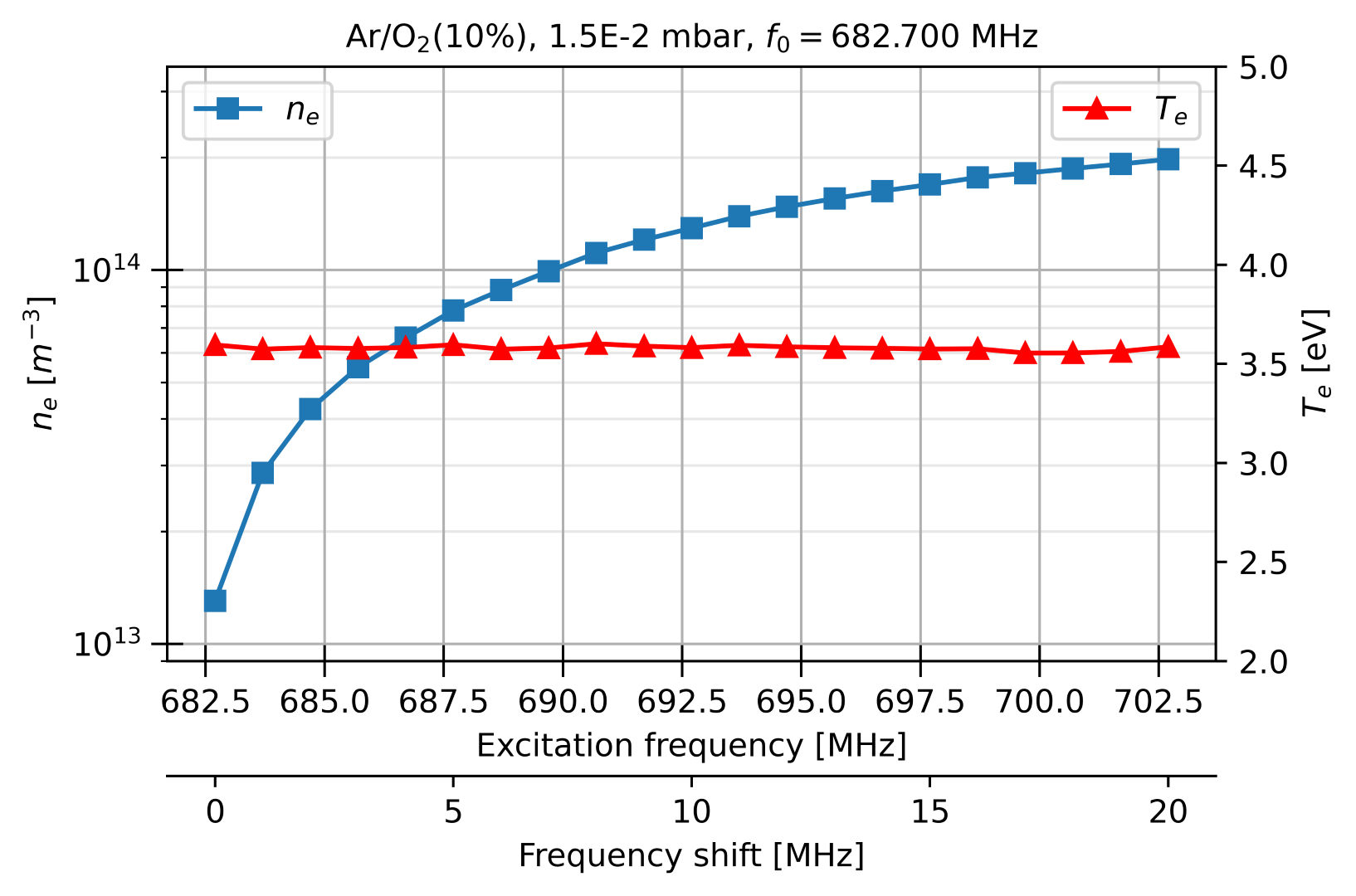}
         \caption{$n_e$ and $T_e$ evolution with frequency shift}
         \label{fig7c}
     \end{subfigure}
	 \caption{Evolution of the EEPF, electron density and electron temperature as a function of the frequency shift for an Ar/O$_2$(10\%) plasma at $1.5\times10^{-2}$ mbar. Vertical lines represent $\varepsilon^*$ and $\varepsilon_i$ being respectively the excitation and ionization threshold of argon.}
	 \label{fig7}
\end{figure}

According to measurements presented in Fig. \ref{fig7c}, electron density is multiplied by a factor 10 with a frequency shift of a few ($\approx10$) MHz. The density increases quickly during the first frequency increments, then tends to steadily increase, or even saturate. This behavior was observed for all the gas mixture investigated: Ar/O$_2$, He/O$_2$, N$_2$/O$_2$ and pure O$_2$. The local electron densities were in the same order of magnitude as well: about $10^{13}$ m$^{-3}$ at ignition and about $10^{14}$ m$^{-3}$ after frequency optimization.

For most of the plasma discharges, $n_e$ is nearly proportional to the power absorbed by the plasma \cite{Godyak_eedf_icp}. An increase of a factor of 10 in density indicates a similar increase in plasma absorbed power, although the RF power was unchanged during the frequency tuning process. 

Fig. \ref{fig7a} and \ref{fig7b} show that EEPFs and EEDFs shape remains almost unchanged across the frequency tuning process. More specifically, Fig. \ref{fig7a} highlights the electron density increase as EEPFs are shifted upward, while Fig. \ref{fig7b} highlights the electron temperature is constant -- although the high-energy tail is slightly less populated at high density (red curve). Even if the electron density increases, the relatively low density does not allow for sufficient electron-electron collisions leading to a ``Maxwellization'' of the EEDF at higher density, as reported in \cite{Godyak_eedf_icp}. 

Finally, frequency tuning does not serve as a control parameter for optimizing cleaning effectiveness, since it does not affect the high-energy tail which controls the plasma chemistry. It only enhances the reactive species production with increasing electron density.

\begin{figure}[htb!]
 \centering
        \includegraphics[width=0.45\textwidth]{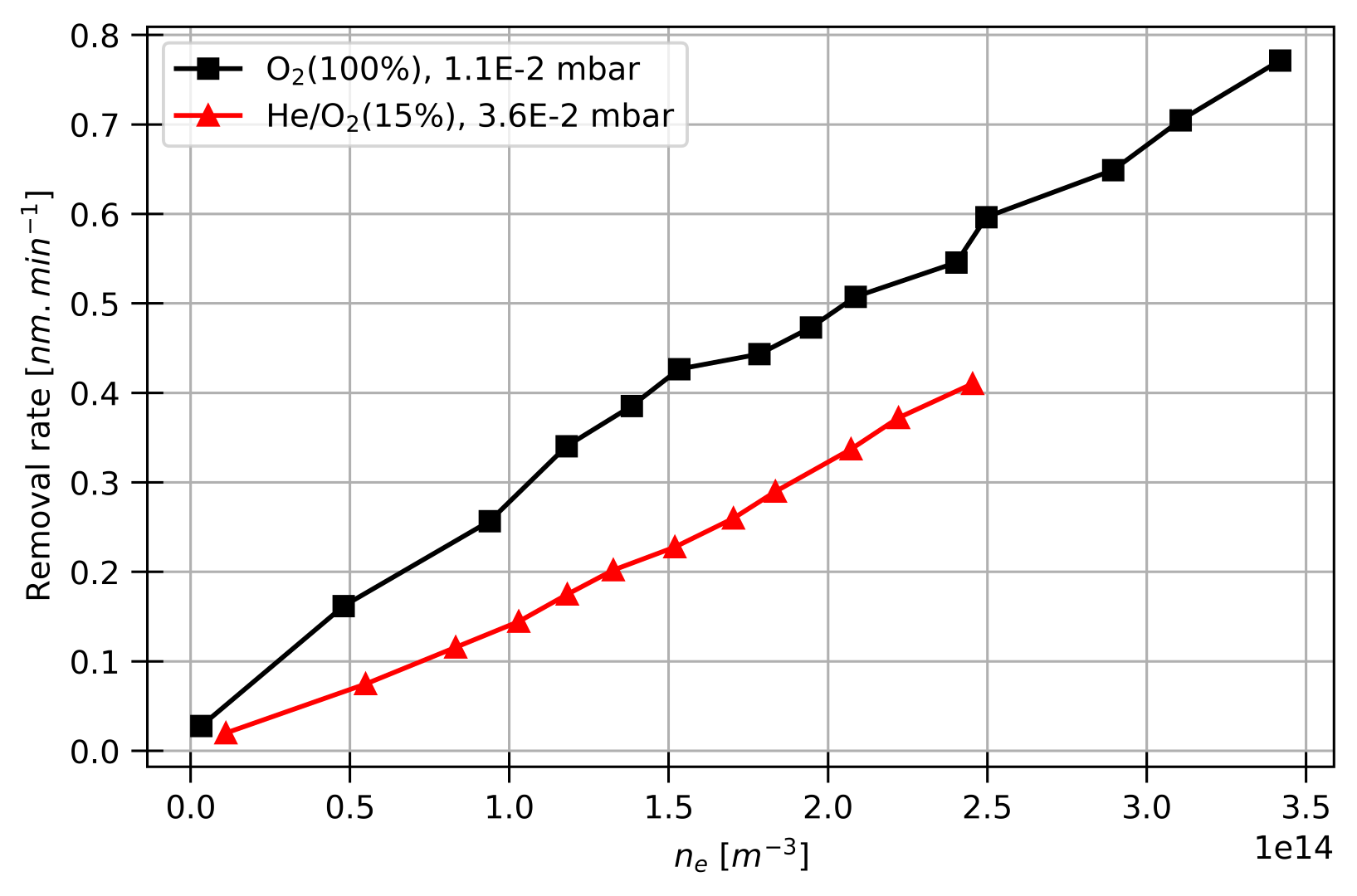}
 \caption{Cleaning rate evolution with electron density for an He/O$_2$(15\%) plasma at $3.6\times10^{-2}$ mbar and a pure O$_2$ plasma at $1.1\times10^{-2}$ mbar. Here, the electron density is controlled by the frequency tuning, as explained earlier.}
\label{fig7bis}
\end{figure}

The cleaning rate is of particular interest because it determines the cleaning effectiveness of a given plasma. The production rate of species in plasma is proportional to $n_e$ (among other parameters). This was verified experimentally for various gas mixtures as shown in Fig. \ref{fig7bis}. A linear evolution of the cleaning rate is observed as a function of electron density. The slope of the curve depends on the gas mixture. 

To summarize, maximizing the electron density is crucial to optimize the production rate of reactive species and thus the cleaning rate. This enhancement can significantly reduce the plasma processing duration of each individual cavity.

\subsection{RF power} \label{section4.4}
\begin{figure}[htb!]
 \centering
        \includegraphics[width=0.45\textwidth]{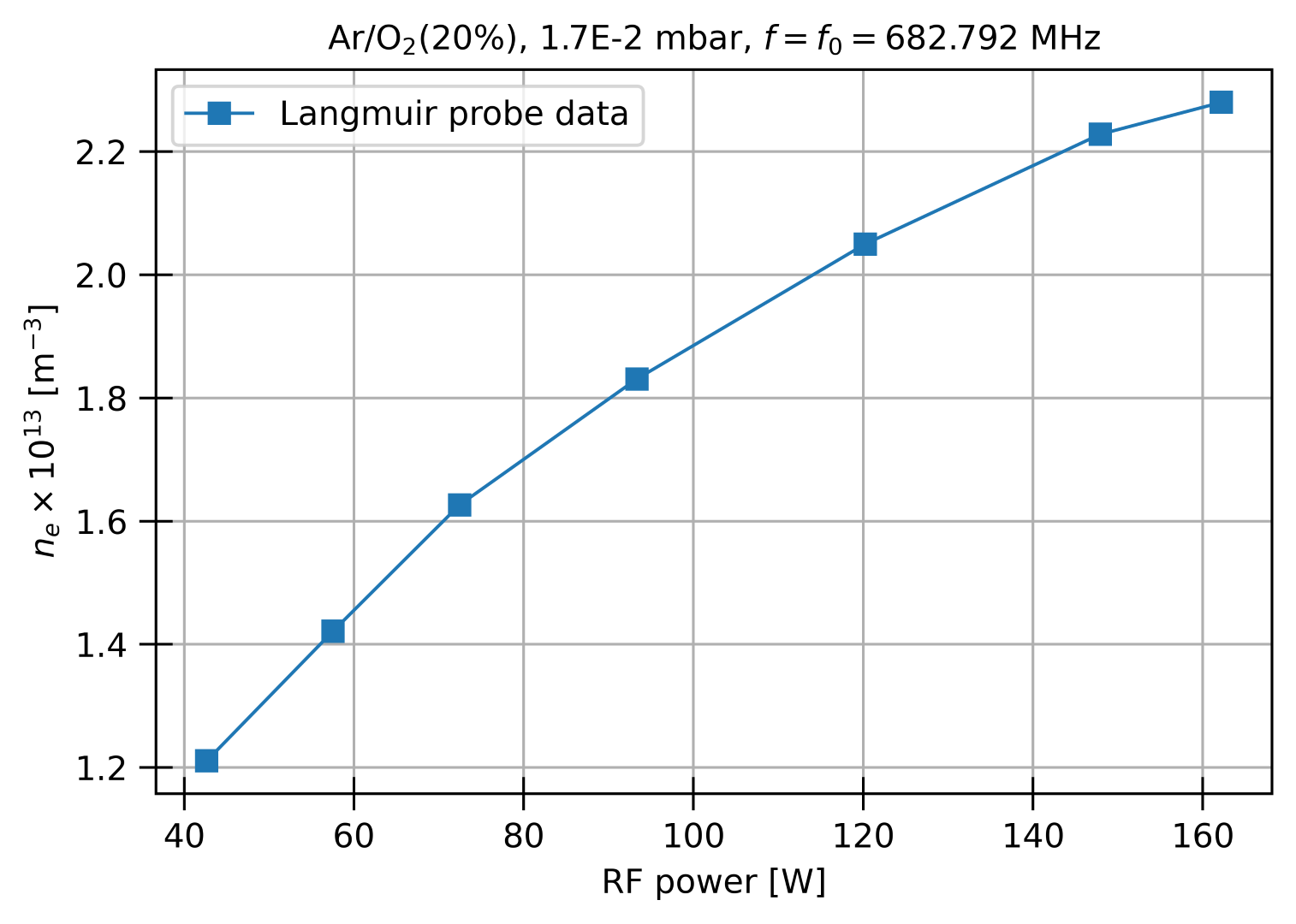}
 \caption{Electron density evolution with the incident RF power for an Ar/O$_2$(20\%) plasma with the drive frequency kept at the resonance under vacuum: $f_0$.}
\label{fig8}
\end{figure}

RF power can be used as well to modulate the electron density -- like the frequency tuning does. However, solely increasing the RF power may lead to the unwanted coupler breakdown and is moreover significantly less effective at enhancing electron density than increasing the excitation frequency, as illustrated in Fig. \ref{fig8}. For instance, increasing the RF power by a factor of 4 only doubles the electron density. In contrast, as shown in Fig. \ref{fig7c}, a 20 MHz frequency shift -- without any change in RF power -- yields a 15 time increase in electron density.

Intuitively, the simultaneous increase of RF power and frequency shift might be expected to enable optimal control of electron density. However, adjustment of the RF power after a frequency shift consistently leads to plasma extinction. The underlying mechanism is not yet fully understood, but the explanation might be that changes in RF power modify the impedance characteristics of the RF transmission line, thereby perturbing the resonance condition and causing plasma extinction.


To summarize, optimal electron density is achieved by first increasing the RF power to ignite the plasma, followed by the frequency tuning. While the RF power can be further increased post-ignition, this approach carries an elevated risk of inducing coupler breakdown. On the other hand, reducing the RF power after ignition sustains the plasma but limits the effectiveness of frequency tuning, as the maximum achievable frequency is lowered. This limitation results in suboptimal electron density, highlighting the trade-off between power and risks management and plasma performance.

\subsection{Gas pressure}
As detailed in Sections \ref{section3.4}, \ref{section4.3}, and \ref{section4.4}, the accessible parameter space for frequency tuning and RF power varies with gas pressure. Consequently, direct comparison of plasma parameters at a fixed RF power and frequency shift across different gas pressures is highly challenging. These constraints arise from the inherent characteristics of the SRF cavity plasma reactor.

\begin{figure}[htb!]
\centering
     \begin{subfigure}[t]{0.32\textwidth}
         \centering
         \includegraphics[width=\textwidth]{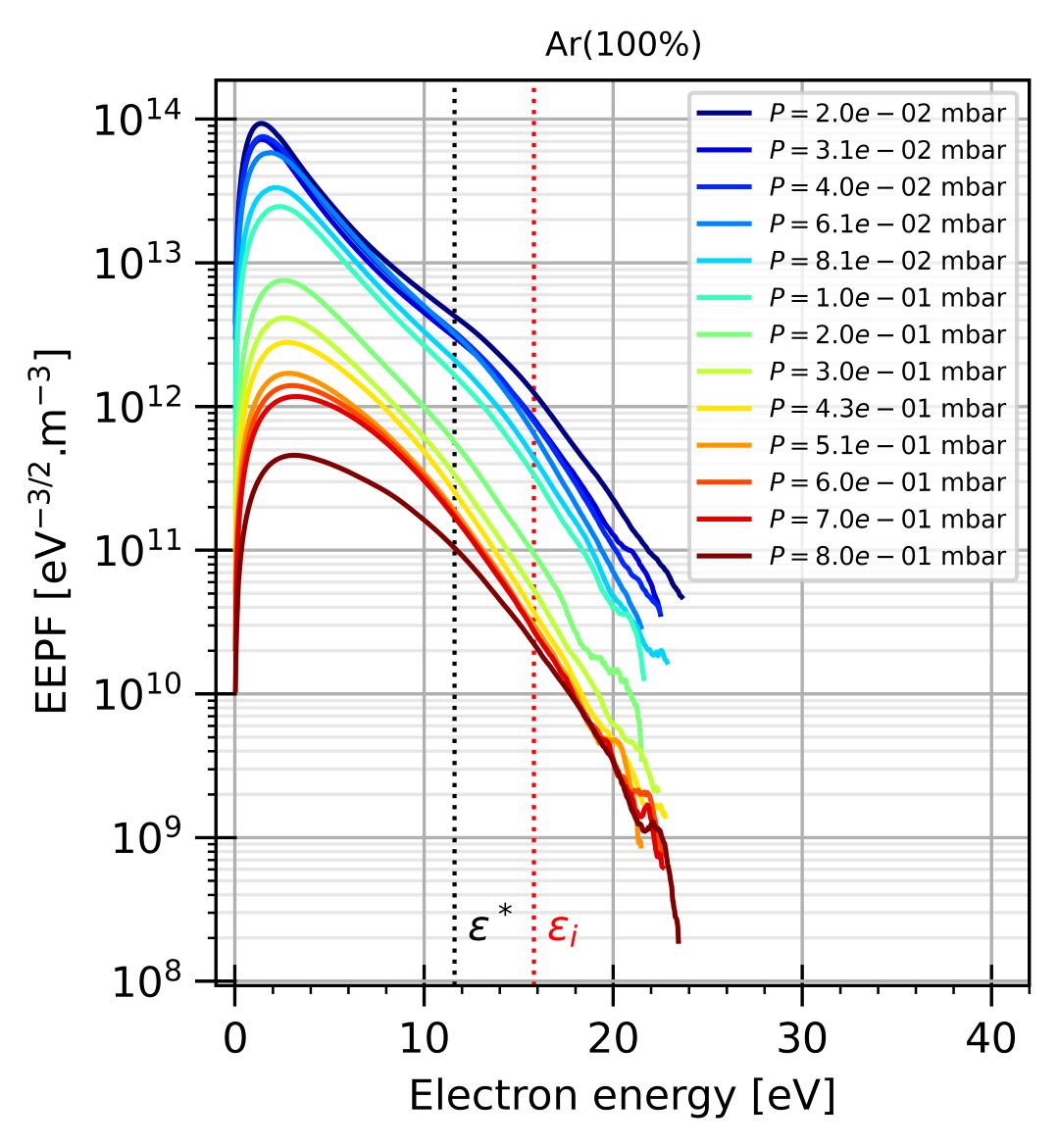}
         \caption{EEPF evolution with gas pressure at the maximum frequency shift}
         \label{fig9a}
     \end{subfigure}
      \hfill
     \begin{subfigure}[t]{0.32\textwidth}
         \centering
         \includegraphics[width=\textwidth]{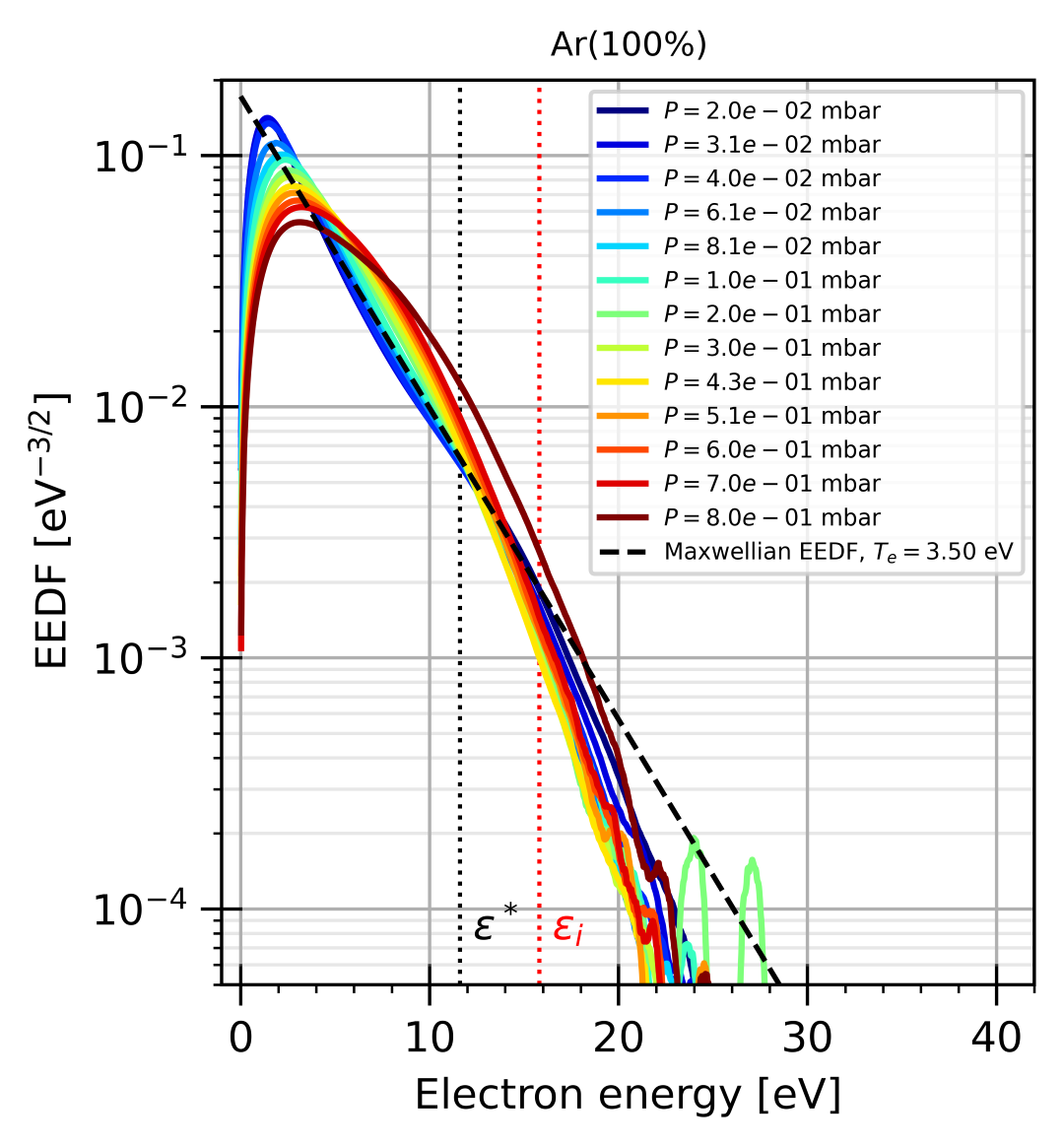}
         \caption{EEDF evolution with gas pressure at the maximum frequency shift}
         \label{fig9b}
     \end{subfigure}
     \hfill
     \begin{subfigure}[t]{0.32\textwidth}
         \centering
         \includegraphics[width=\textwidth]{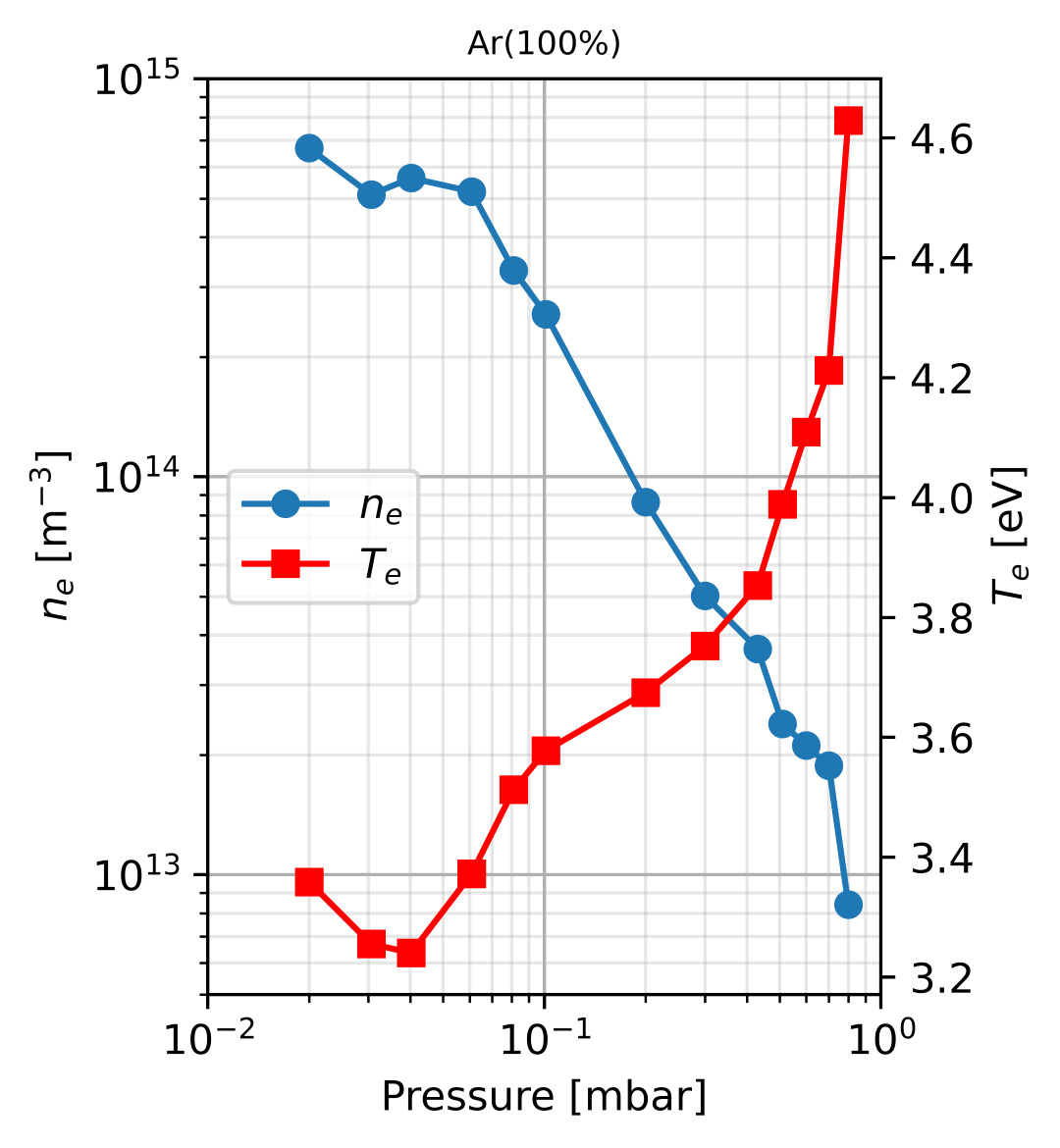}
         \caption{$n_e$ and $T_e$ evolution with gas pressure at the maximum frequenyc shift}
         \label{fig9c}
     \end{subfigure}
	 \caption{Evolution of the EEPF, EEDF, electron density and electron temperature as a function of the gas pressure for an Ar plasma at the maximum frequency shift. Vertical lines represent $\varepsilon^*$ and $\varepsilon_i$ being respectively the excitation and ionization threshold of argon.}
	 \label{fig9}
\end{figure}

To investigate the pressure dependence of plasma parameters, the measurement point was defined according to the following criteria: at each pressure, the RF power was set to the minimum value required for plasma ignition. With this power held constant, the frequency was then shifted to its maximum value before plasma extinction (or mode transition in the low-pressure domain). The measurements of plasma parameters respecting these criteria are given in Fig. \ref{fig9} for a pure argon plasma in the pressure range $[2\times10^{-2}~-~8\times10^{-1}]$ mbar.

Figure \ref{fig9c} shows a pronounced decrease in electron density with increasing pressure. A reduction of nearly two orders of magnitude is observed between $2\times10^{-2}$ mbar and $8\times10^{-1}$ mbar, with $n_e$ declining from $6.7\times10^{14}$ to $8.3\times10^{12}$ m$^{-3}$. This trend is primarily attributed to some technical limitations and more specifically the limited frequency optimization achievable at high pressures. A plateau in electron density is observed between $2\times10^{-2}$ and $6\times10^{-2}$ mbar, as the maximum frequency shift remains relatively constant within this range. Beyond $6\times10^{-2}$ mbar, however, the maximum achievable frequency shift diminishes, resulting in the observed reduction in electron density. 


On the contrary, the electron temperature is increasing as the pressure increases. 
At low pressure, the EEDF is a three-temperature-like distribution with a pronounced low-energy bump, while at higher pressure, it evolves to a Druyvesteyn-like EEDF due to more collisions with the neutral gas. It can be seen on Fig. \ref{fig9c} that over the whole pressure range investigated, the high-energy tail ($\varepsilon>\varepsilon_i$) temperature is constant with $T_{e,high}\approx 8.5$ eV. Indeed, as the pressure increases, the low energy population is shifted to medium energy, while the high-energy tail remains unaffected.

The drop in electron density in Fig. \ref{fig9c} with increasing pressure is observed for all the gas mixtures investigated for the same reason: limited frequency tuning range. This reduction in electron density directly impacts the plasma's cleaning capabilities, a relationship quantified using the installed carbon-coated quartz crystal microbalance (QCM).

\begin{figure}[htb!]
 \centering
        \includegraphics[width=0.4\textwidth]{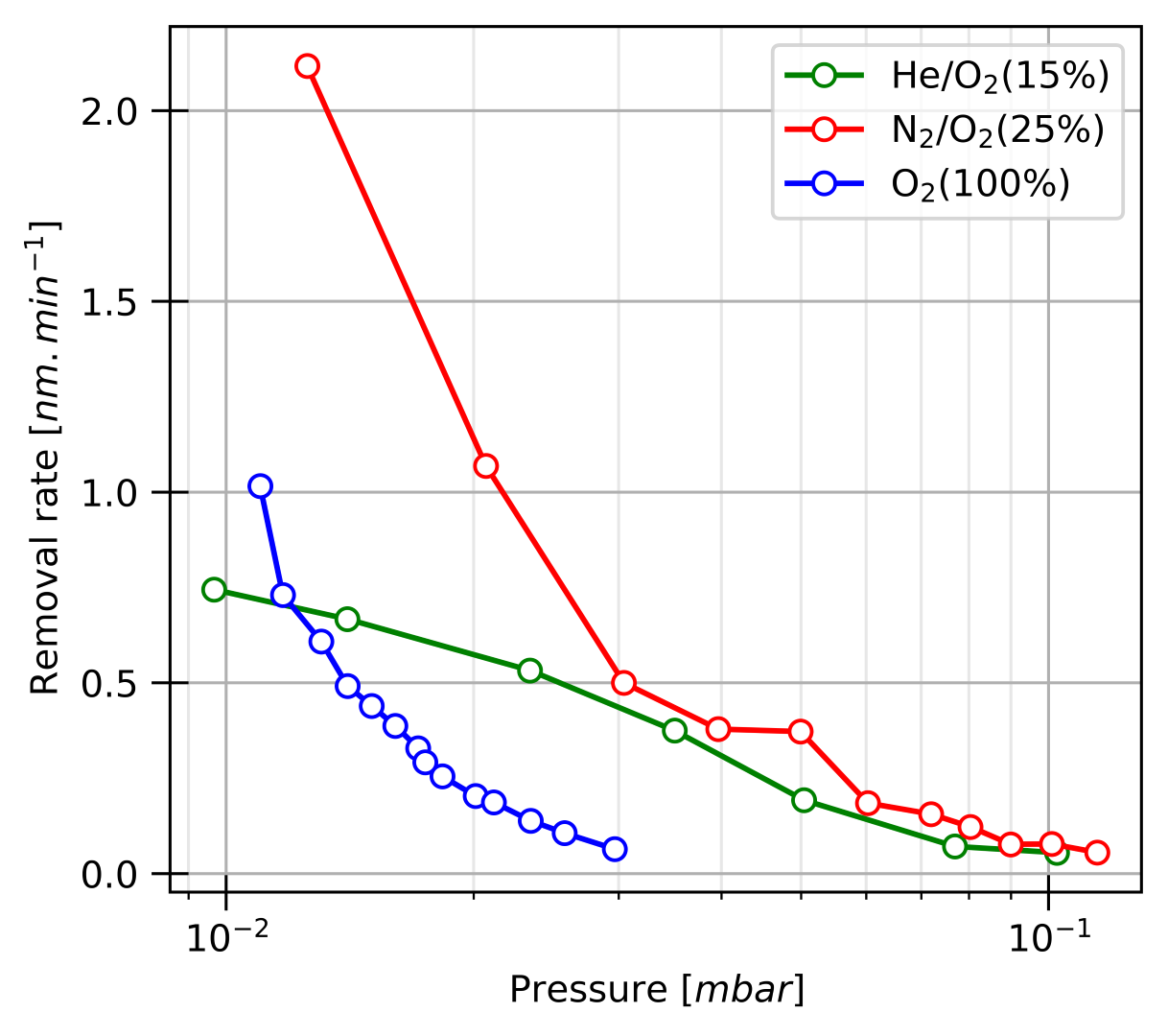}
 \caption{Amorphous carbon cleaning rate variation with gas pressure for various gas mixtures.}
\label{fig122}
\end{figure}

Beyond its effect on electron density, pressure also plays a critical role in governing collision and diffusion processes within the plasma. At low pressures, the increased mean free path of neutral particles reduces recombination and quenching of reactive species, thereby enhancing cleaning effectiveness. This behavior is clearly illustrated in Fig.~\ref{fig122}, where a decline in removal rate is observed with increasing pressure across all investigated gas mixtures. 

The carbon removal rate monotonically approaches zero as pressure rises, regardless of the gas type. However, the rate of this decrease varies between mixtures, as the EEDF governing plasma chemistry is differentially affected by pressure changes. This mixture-dependent behavior arises because each species exhibits distinct excitation and ionization thresholds, cross sections, excited state energies, reactive species mean free paths, and recombination and quenching rates. Consequently, each gas mixture undergoes unique modifications in both its EEDF and cleaning efficiency.

\subsection{Gas composition}
Several gas mixtures were tested with simultaneous Langmuir probe and cleaning rate measurements. Argon-oxygen, helium-oxygen, nitrogen-oxygen with varying oxygen concentration (0 to $\approx 40$\%) and pure oxygen have been investigated. Plasma parameters and cleaning rates as a function of gas composition are discussed in this section.


\subsubsection{Oxygen concentration, dilution gas \& plasma parameters}
In this subsection, the effect of the oxygen concentration on the plasma parameters are discussed for an Ar/O$_2$ along with an He/O$_2$ mixture.

\begin{figure}[htb!]
\centering
     \begin{subfigure}[t]{0.5\textwidth}
         \centering
         \includegraphics[width=\textwidth]{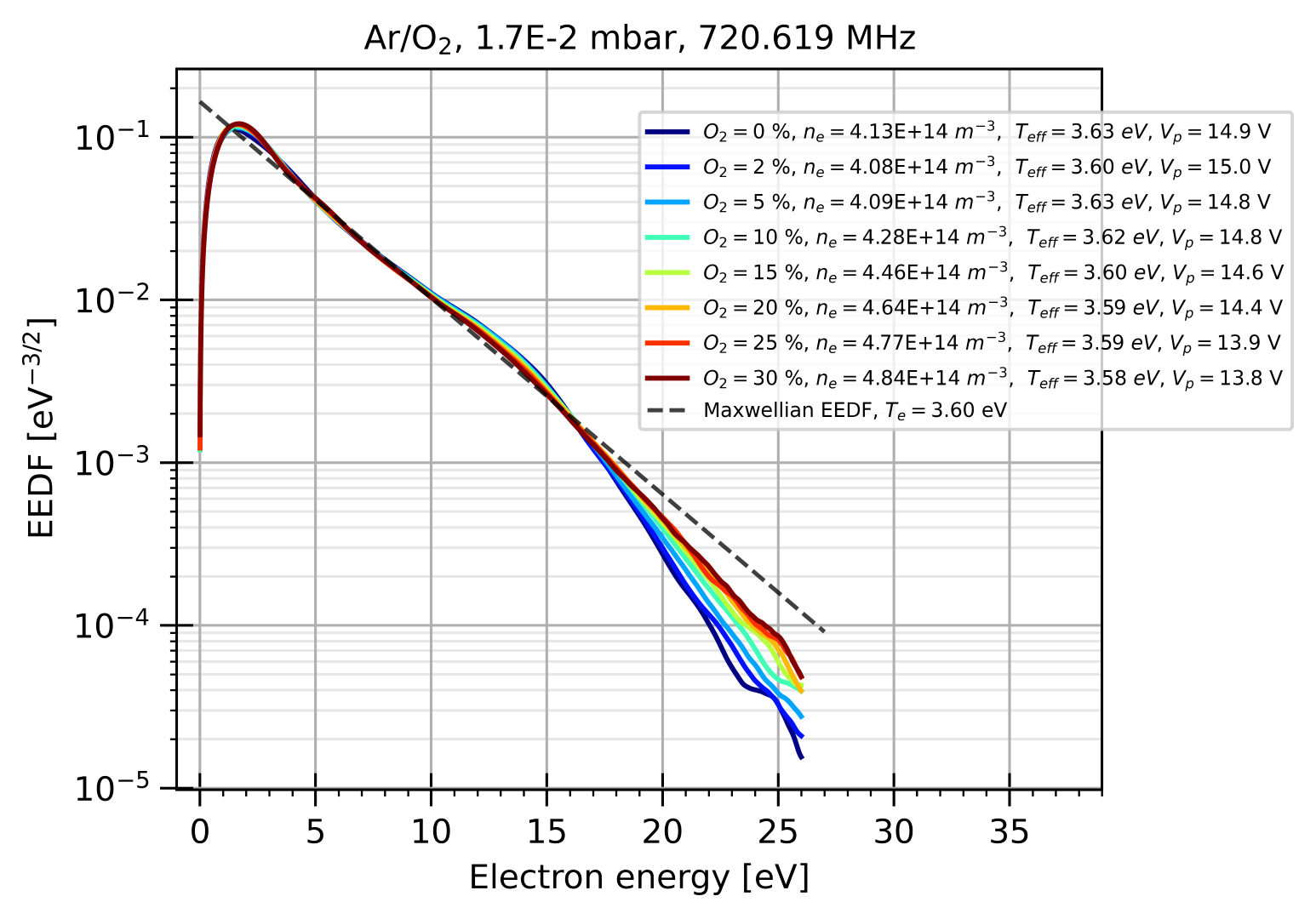}
         \caption{Measured EEDF for different O$_2$ concentration in Ar at a fixed pressure of $1.7\times10^{-2}$ mbar and $\Delta f=38$ MHz. Vertical lines represent $\varepsilon^*$ and $\varepsilon_i$ of argon.}
         \label{fig13a}
     \end{subfigure}
     \hfill
     \begin{subfigure}[t]{0.45\textwidth}
         \centering
         \includegraphics[width=\textwidth]{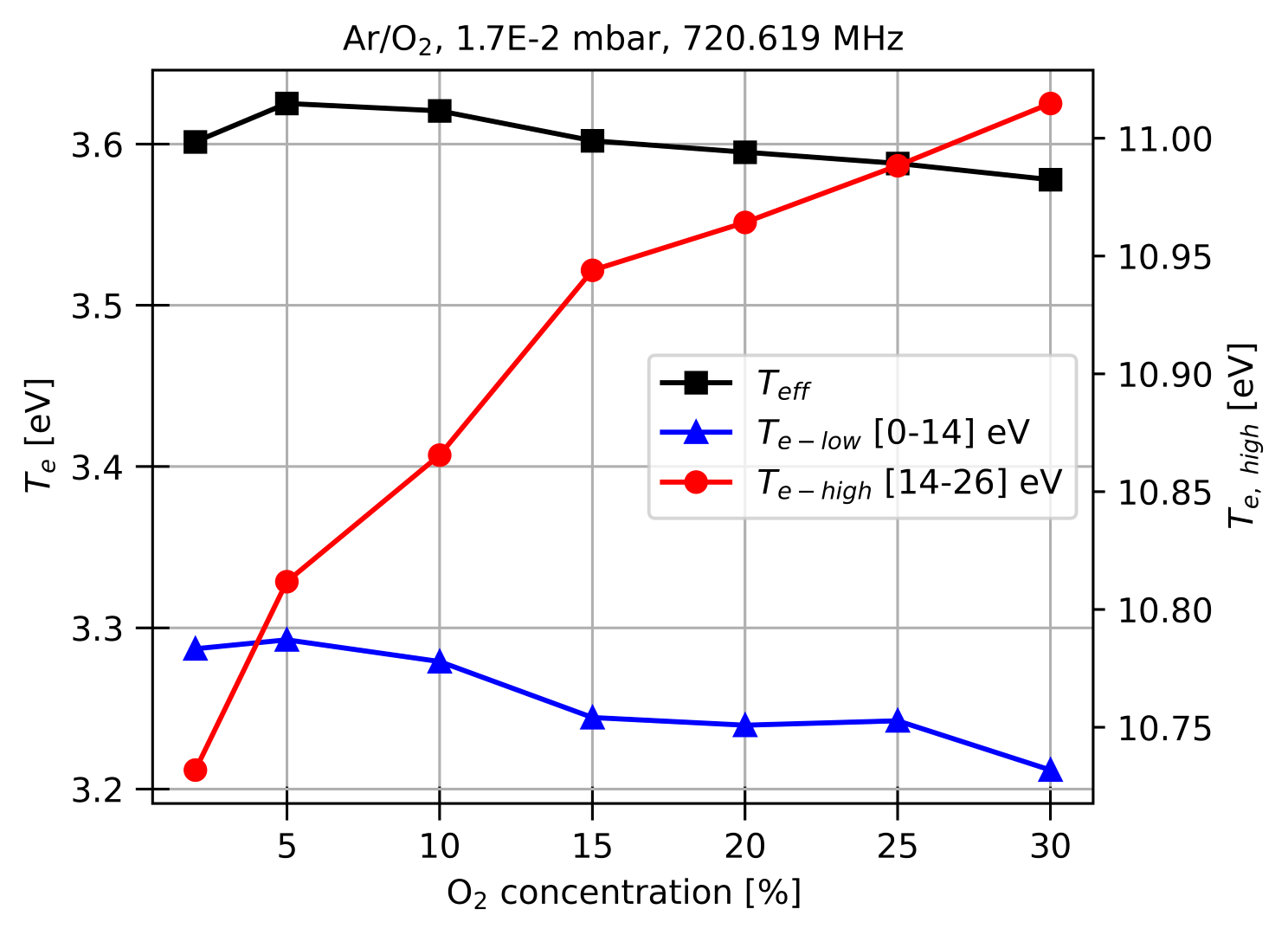}
	 \caption{Measured $T_{eff}$, $T_{e-low}$ and $T_{e-high}$ for different O$_2$ concentration in Ar at a fixed pressure of $1.7\times10^{-2}$ mbar and $\Delta f=38$ MHz. $T_{e-high}$ is represented on the secondary y-axis.}
         \label{fig13b}
     \end{subfigure}
	 \caption{Measured EEDF and $T_{e}$ for different O$_2$ concentration in Ar at a fixed pressure of $1.7\times10^{-2}$ mbar and $\Delta f=38$ MHz.}
	 \label{fig13}
\end{figure}

Fig. \ref{fig13a} shows the electron energy distribution functions (EEDFs) for an argon-oxygen plasma at $1.7\times10^{-2}$ mbar and a frequency shift of $\Delta f=38$~MHz, for oxygen concentrations from 0\% to 30\%. The oxygen fraction was varied while maintaining the plasma ignited. The low-energy portion of the EEDF (below 14 eV) remains largely unchanged across these concentrations, while the high-energy tail evolves, showing an increase in the tail temperature with rising oxygen content. This trend is quantified in Fig.~\ref{fig13b}, where $T_{e,high}$ rises from 10.73 eV at 0\% O$_2$ to 11.01 eV at 30\% O$_2$. Meanwhile, the electron density $n_e$ increases modestly from $4.13\times10^{14}$~m$^{-3}$ to $4.84\times10^{14}$~m$^{-3}$.

Under these specific conditions (low pressure, fixed power and frequency shift), the addition of oxygen introduces strong inelastic loss channels -- including vibrational and electronic excitation, dissociation, and electron attachment leading to negative ions (O$^-$ and O$_2^-$) -- which could deplete electron density and energy. However, the observed slight increase in $n_e$ suggests that enhanced ionization from the high-energy tail (via Ar at 15.8~eV threshold, O$_2$ at 13.6~eV, and O at 12~eV) tends to compensate electron losses due to attachment processes, allowing the discharge to self-adjust. The low-energy range remains similar because it is primarily shaped by elastic collisions and bulk heating processes that are less impacted by the oxygen admixture in this regime.

\begin{figure}[htb!]
\centering
     \begin{subfigure}[t]{0.45\textwidth}
         \centering
         \includegraphics[width=\textwidth]{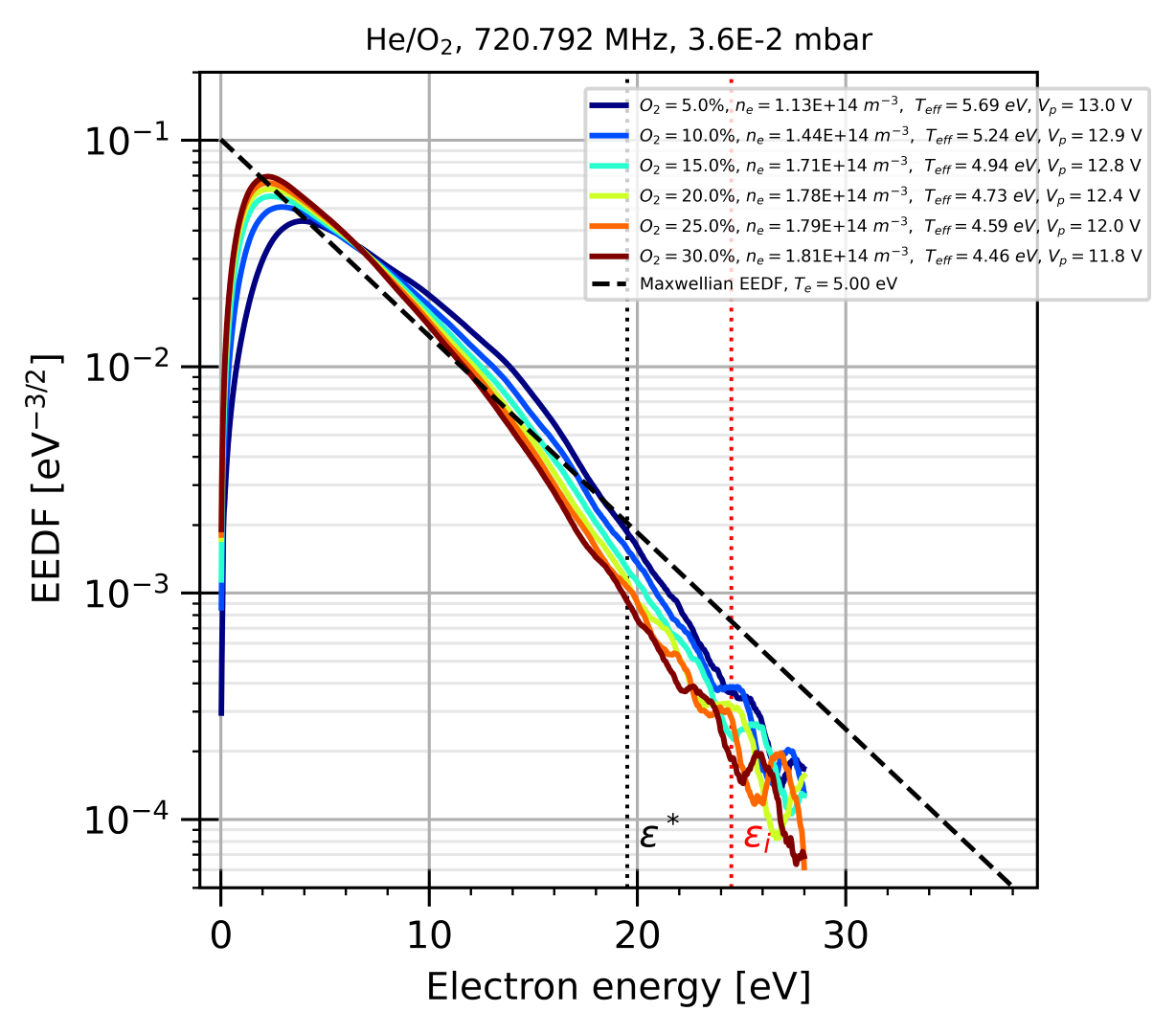}
         \caption{Measured EEDF for different O$_2$ concentration in He at a fixed pressure of $3.6\times10^{-2}$ mbar and $\Delta f=38$ MHz. Vertical lines represent $\varepsilon^*$ and $\varepsilon_i$ of argon.}
         \label{fig14a_}
     \end{subfigure}
     \hfill
     \begin{subfigure}[t]{0.45\textwidth}
         \centering
         \includegraphics[width=\textwidth]{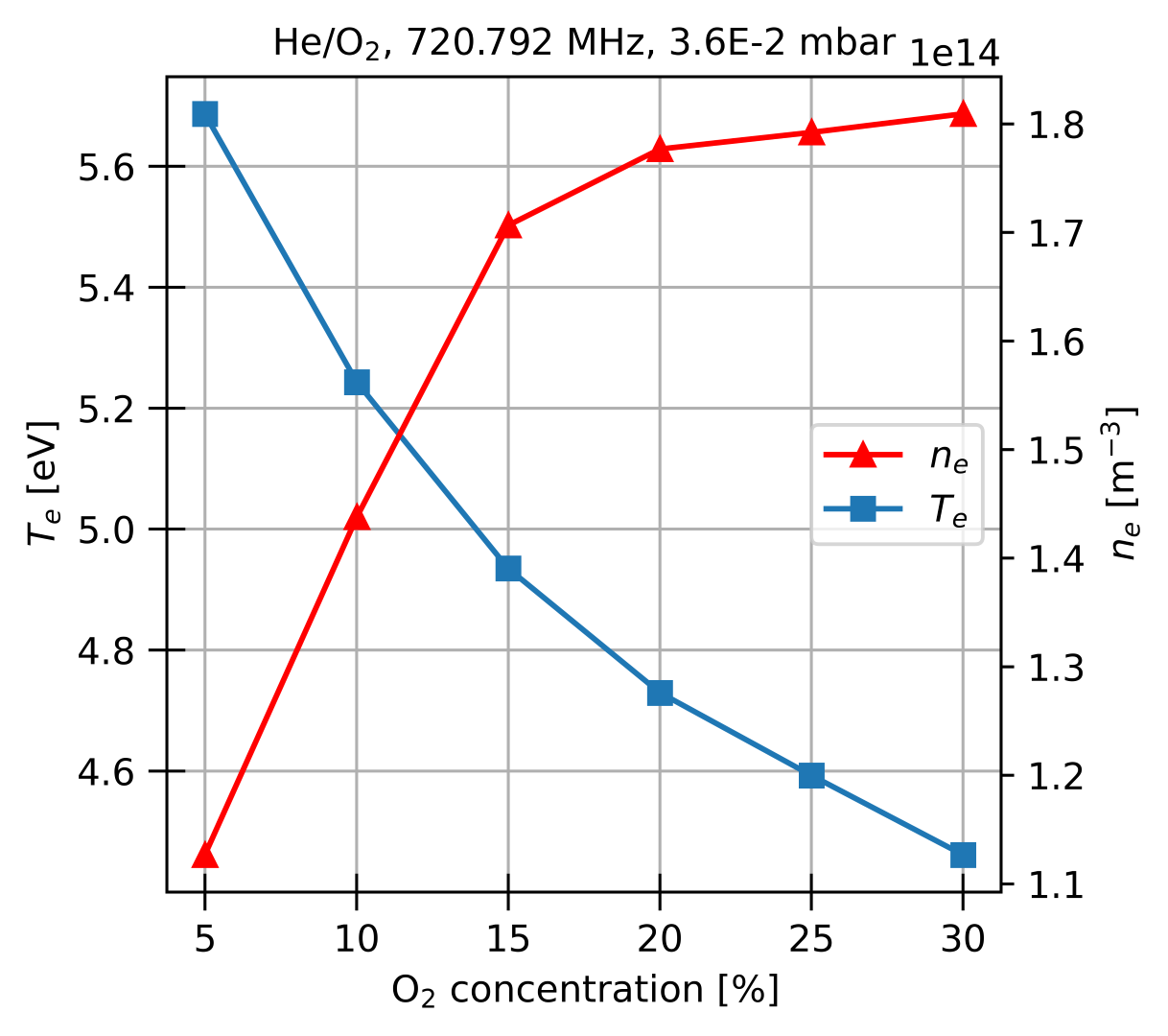}
	 \caption{Measured $T_{eff}$, $n_e$ for different O$_2$ concentration in He at a fixed pressure of $3.6\times10^{-2}$ mbar and $\Delta f=38$ MHz.}
         \label{fig14b_}
     \end{subfigure}
	 \caption{Measured EEDF and $T_{e}$ for different O$_2$ concentration in He at a fixed pressure of $3.6\times10^{-2}$ mbar and $\Delta f=38$ MHz.}
	 \label{fig14_}
\end{figure}

Fig. \ref{fig14a_} shows the EEDFs for a helium–oxygen plasma at $3.6\times10^{-2}$ mbar and a frequency shift of $\Delta f = 38$ MHz, for oxygen concentrations from 5\% to 30\%. The oxygen fraction was varied while maintaining the plasma discharge, as in the argon case. Under these operating conditions, the plasma systematically extinguished when the oxygen content exceeded 35-40\%. This is attributed to the depletion of helium metastables (He$^*$) by O$_2$ through Penning ionization  (He\(^*\) + O\(_2\) \(\rightarrow\) O\(_2^+\) + He + e\(^-\))  and related collisional processes. Given He metastables normally sustain the discharge  by enabling stepwise ionization (He\(^*\) + e\(^-\) \(\rightarrow\) He\(^+\) + 2e\(^-\)), which lowers the effective ionization threshold. The drain of He$^*$ by O$_2$ thus disrupts this mechanism, leading the discharge to quench.

Unlike the Ar/O$_2$ case (Fig.~\ref{fig13a}), where primarily the high-energy tail heated up (with $T_{e,\mathrm{high}}$ increasing from 10.7~eV to 11~eV), the He/O$_2$ EEDF shows modification across the full distribution with added oxygen, reflecting He plasmas’ higher sensitivity to inelastic losses owing to elevated excitation/ionization thresholds and metastable reliance. This is evident in Fig.~\ref{fig14b_}, where the bulk effective electron temperature $T_e$ gradually decreases with increasing oxygen content, despite a 60\% rise in electron density between 5\% and 20\% O$_2$.

The increase in \(n_e\) arises because Penning ionization (He\(^*\) + O\(_2\) \(\rightarrow\) O\(_2^+\) + He + e\(^-\)) and lower-threshold ionization of O\(_2\)/O (ionization energies 12/13.6~eV, versus 24.6~eV for He) initially enhance electron production.

\begin{figure}[htb!]
 \centering
        \includegraphics[width=0.4\textwidth]{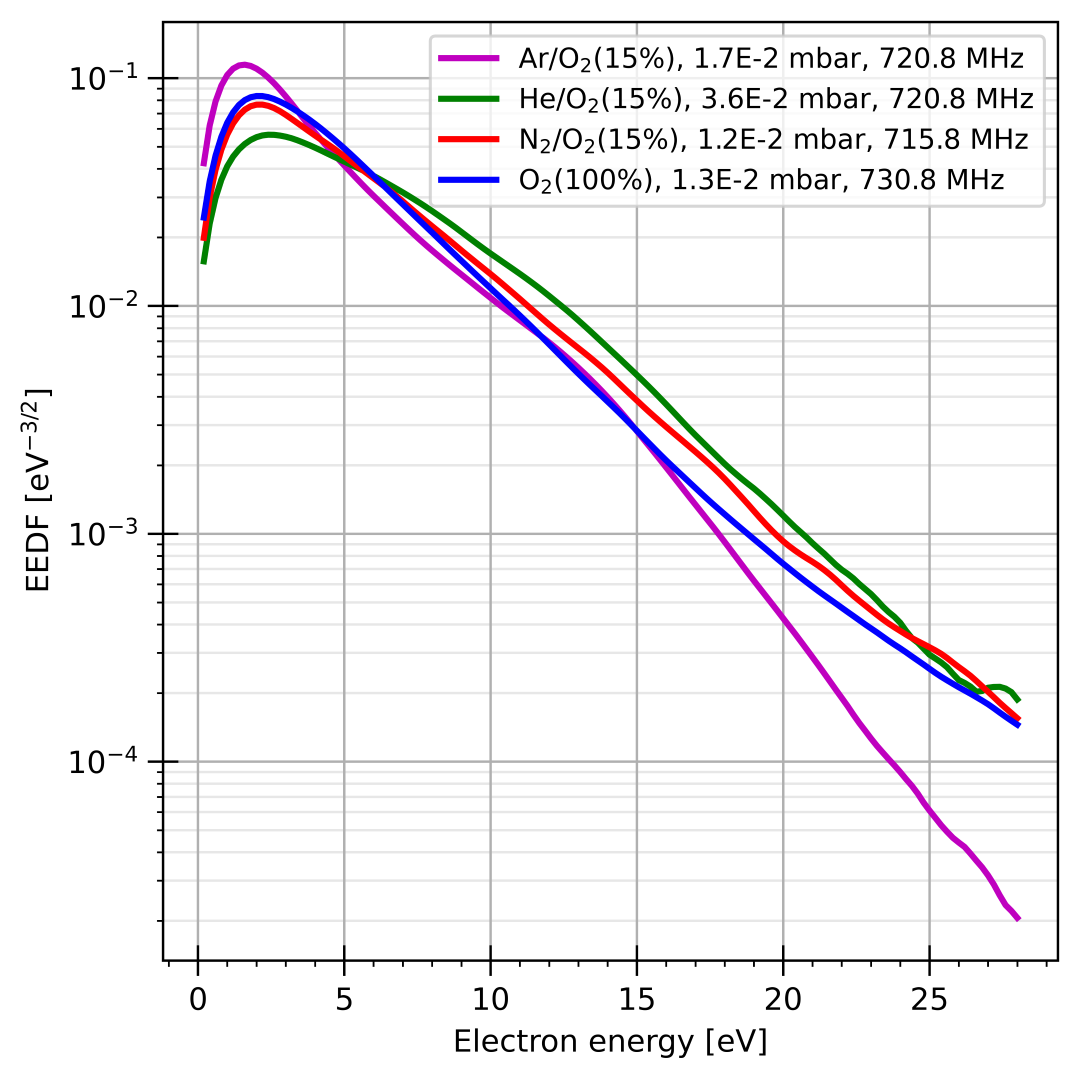}
 \caption{EEDF for various gas mixture in similar experimental conditions.}
\label{fig14}
\end{figure}

Fig. \ref{fig14} represents EEDFs for the investigated gas mixtures in similar experimental conditions. The distribution functions are relatively similar for Ar/O$_2$, N$_2$/O$_2$ and pure O$_2$ with an almost Maxwellian shape in the energy range below 15 eV. However, in the case of He/O$_2$, the distribution is convex. This is attributed to the higher pressure operating point for helium-oxygen plasma. In the energy range $>15$ eV, the high energy tail is depleted for Ar/O$_2$ mixture, while this depletion effect is much less pronounced for He/O$_2$, N$_2$/O$_2$ and pure O$_2$ mixtures. These differences in EEDF shape are expected to impact the cleaning rates as it controls the plasma chemistry. This is illustrated in Fig. \ref{fig15} where cleaning rates are measured for various dilution gases and for various O$_2$ concentration.

\subsubsection{Oxygen concentration, dilution gas \& carbon cleaning rates}
In this subsection, the effect of the oxygen concentration on the carbon cleaning rates are discussed for several oxygen-based mixtures.

\begin{figure}[htb!]
 \centering
        \includegraphics[width=0.4\textwidth]{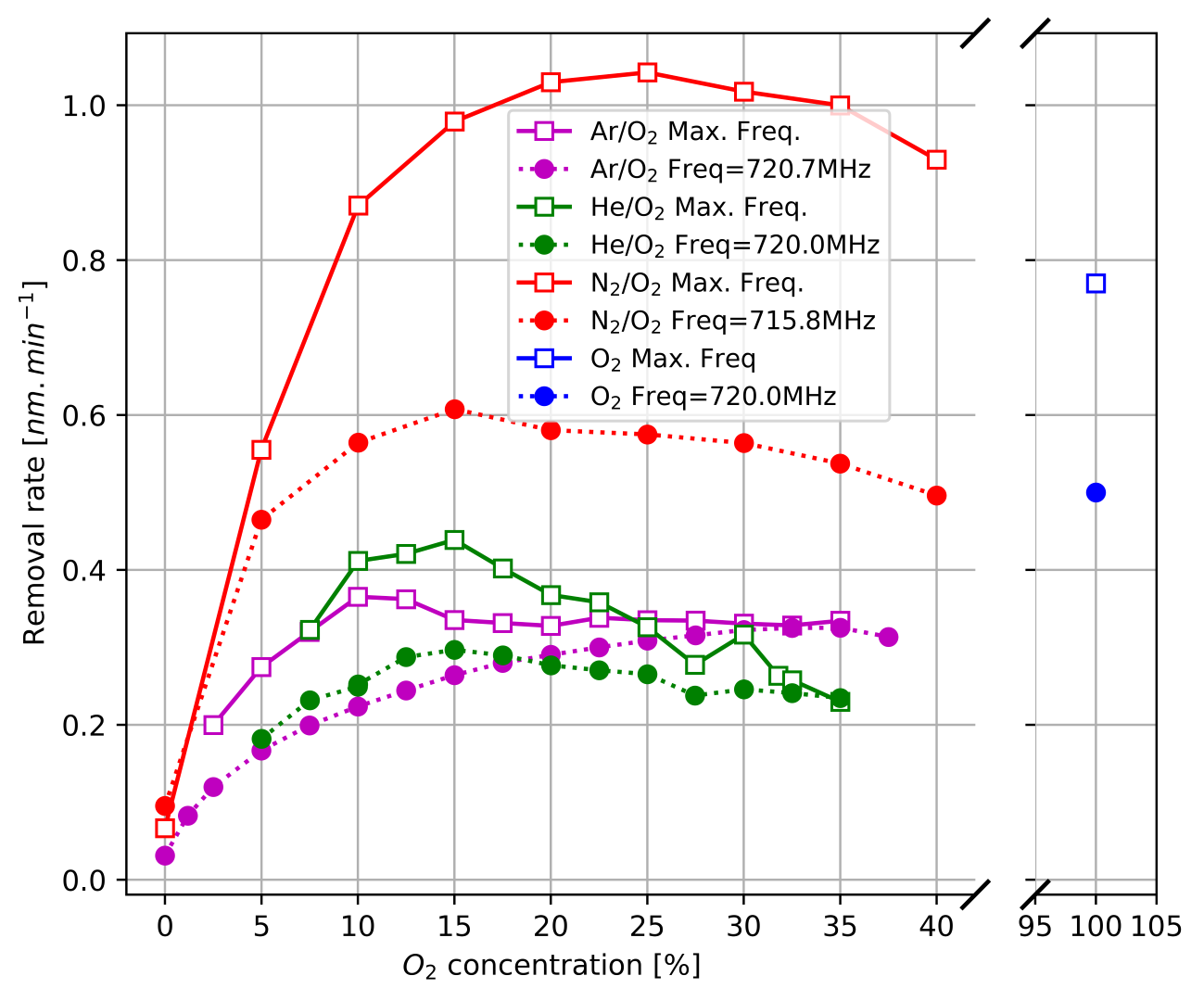}
 \caption{Cleaning rates for different dilution gases as a function of oxygen concentration. Square markers represent measurements at the highest frequency shift attainable with each mixture (``Max. Freq.''), while dot markers correspond to measurements at a constant frequency shift as the O$_2$ concentration is increased, the oxygen fraction being in this case the only parameter varied. The total gas pressure is $1.8\times10^{-2}$ mbar for Ar/O$_2$, $3.7\times10^{-2}$ mbar for He/O$_2$, and $1.3\times10^{-2}$ mbar for both N$_2$/O$_2$ and pure O$_2$.}
\label{fig15}
\end{figure}

Figure \ref{fig15} shows the dependence of the cleaning rate on the oxygen concentration for Ar/O$_2$, He/O$_2$, N$_2$/O$_2$, and pure O$_2$ plasmas. Measurements were performed under two distinct operating conditions. In the first case (labeled ``Max. Freq.''), the cleaning rate was measured at the highest attainable frequency shift for each oxygen concentration, which corresponds to the condition of maximum electron density and thus approximates the upper limit of the cleaning performance for that composition. In the second case (labeled ``Freq = XXX.X MHz''), the cleaning rate was recorded at a fixed frequency shift while varying the O$_2$ concentration from 0\% to 40\%, in order to isolate the influence of oxygen content on the cleaning rate under nominally constant plasma excitation conditions.

In the first scenario (square markers), each dilution gas exhibits a maximum cleaning rate at a specific oxygen content: approximately 10\% O$_2$ for argon, 15\% for helium, and 25\% for nitrogen. These maxima coincide with the oxygen fractions at which the maximum frequency shift, and therefore the highest electron density, is obtained for the corresponding mixture. This indicates that, under conditions of maximized excitation, the optimal cleaning rate is closely linked to the mixture composition that maximizes electron density and, consequently, the production reactive species.

In the second scenario (dot markers), the electron density is initially constrained by the fixed frequency condition, and the peak cleaning rates occur at different oxygen concentrations than in the first case. For Ar/O$_2$, the maximum cleaning rate is observed around 35\% O$_2$, whereas for both He/O$_2$ and N$_2$/O$_2$ it appears near 15\% O$_2$. For all dilution gases, the cleaning rate as a function of oxygen fraction displays a bell-shaped behavior: it initially increases with O$_2$ content, due to the enhanced production of reactive oxygen species, and then decreases beyond a certain threshold.  The decrease can be attributed to a modification of the electron energy distribution function that shifts electrons towards lower energies, as shown in Fig. \ref{fig14_} for He/O$_2$, reducing oxygen reactive species production. As the study is performed at constant pressure, the decrease in cleaning rate can only be attributed to the production rate of reactive species and not to recombination/quenching attributed to collisions in the case of a pressure dependent study.

Since the oxygen fraction at which the maximum achievable electron density is obtained does not necessarily coincide with the fraction at which the cleaning efficiency is highest under fixed excitation conditions, a compromise operating point must be identified. In the case of He/O$_2$, an oxygen content of about 15\% appears optimal in both scenarios, suggesting that this composition provides a favorable balance between high electron density and efficient production of reactive oxygen species for plasma cleaning.

\subsubsection{Reaction rates \& carbon cleaning rates}
In this subsection, the reactive species reaction rates are computed and correlated with the carbon cleaning rates.

From the EEDF, the electron impact reaction rates can be computed with the following equations:

\begin{equation} \label{eq:3}
k_k=\sqrt{\frac{2e}{m_e}}\int_{\varepsilon_{thr}}^{+\infty} \varepsilon \sigma_k F_0 d\varepsilon ~~[m^3.s^{-1}]
\end{equation}

\begin{equation} \label{eq:4}
R_k= k_k x_k N n_e ~~[m^{-3}.s^{-1}]
\end{equation}

Where $\varepsilon_{thr}$ is the threshold energy of the k-th electron impact reaction, $\sigma_k$ is the cross section of the k-th reaction, $F_0$ is the normalized EEDF defined as $\int_0^{+\infty}\varepsilon^{1/2} F_0 d\varepsilon=1$, $x_k$ is the molar fraction of the k-th specie, $N$ is the neutral gas density and $n_e$ the electron density.

The reaction rates for the He/O$_2$ plasma operated at a pressure of $3.6\times10^{-2}$~mbar and a frequency shift of 38~MHz were calculated as a function of the oxygen concentration using the EEDFs presented in Fig.~\ref{fig14_}. Electron-impact cross sections were taken from the Biagi database available through LXCat \cite{Biagi_LXCat}. The associated reactions used for reaction rates calculation are summarized in Table \ref{table1} for O$_2$ and in Table \ref{table2} for He.

\begin{table}[htb!]
\centering
\caption{List of electron impact reaction for O$_2$. Data taken from the Biagi database available through LXCat \cite{Biagi_LXCat}.}
\begin{tabular}{c|c|c|c}
\textbf{Reaction} & \textbf{Formula}                                 & \textbf{Type}                     & \textbf{$\Delta\varepsilon$ {[}eV{]}} \\ \hline
R1       & $e + O_2 \rightarrow O^-+O$                           & Dissociative attachment   & --                                        \\
R2       & $e + O_2 \rightarrow e + O_2$                          & Elastic                               & --                                         \\
R3       & $e + O_2 \rightarrow e + O_2$                          & Excitation, rotation           & 0.02                                     \\
R4       & $e + O_2 \rightarrow e + O_2$                          & Excitation, vibration         & 0.193                                    \\
R5       & $e + O_2 \rightarrow e + O_2$                          & Excitation, vibration         & 0.386                                    \\
R6       & $e + O_2 \rightarrow e + O_2$                          & Excitation, vibration         & 0.579                                    \\
R7       & $e + O_2 \rightarrow e + O_2$                          & Excitation, vibration         & 0.772                                    \\
R8       & $e + O_2 \rightarrow e + O_2$                          & Excitation, vibration         & 0.977                                     \\
R9       & $e + O_2 \rightarrow e + O_2$                          & Excitation, vibration         & 1.627                                     \\
R10     & $e + O_2 \rightarrow e + O_2$                          & Excitation, vibration         & 4.5                                         \\
R11     & $e + O_2 \rightarrow e + O + O$                      & Excitation, dissociation    & 6.1                                          \\
R12     & $e + O_2 \rightarrow e + O + O$                      & Excitation, dissociation    & 8.4                                           \\
R13     & $e + O_2 \rightarrow e + O_2$                          & Excitation                         & 9.3                                           \\
R14     & $e + O_2 \rightarrow e + e + {O_2}^+$            & Ionization                         & 12.072             
\end{tabular}
\label{table1}
\end{table}

\begin{table}[htb!]
\centering
\caption{List of electron impact reaction for He. Data taken from the Biagi database available through LXCat \cite{Biagi_LXCat}.}
\begin{tabular}{c|c|c|c}
\textbf{Reaction}  & \textbf{Formula}                               & \textbf{Type}    & \textbf{$\Delta\varepsilon$ {[}eV{]}} \\ \hline
R15                      & $e + He \rightarrow e + He$            & Elastic              & --                                        \\
R16                      & $e + He \rightarrow e+ He(2^3S)$  & Excitation        & 19.82                                   \\
R17                      & $e + He \rightarrow e +He(2^1S) $ & Excitation        & 20.61                                   \\
R18                      & $e + He \rightarrow e + He^+$       & Ionization        & 24.587               
\end{tabular}
\label{table2}
\end{table}

\begin{figure}[htb!]
\centering
     \begin{subfigure}[t]{0.49\textwidth}
         \centering
         \includegraphics[width=\textwidth]{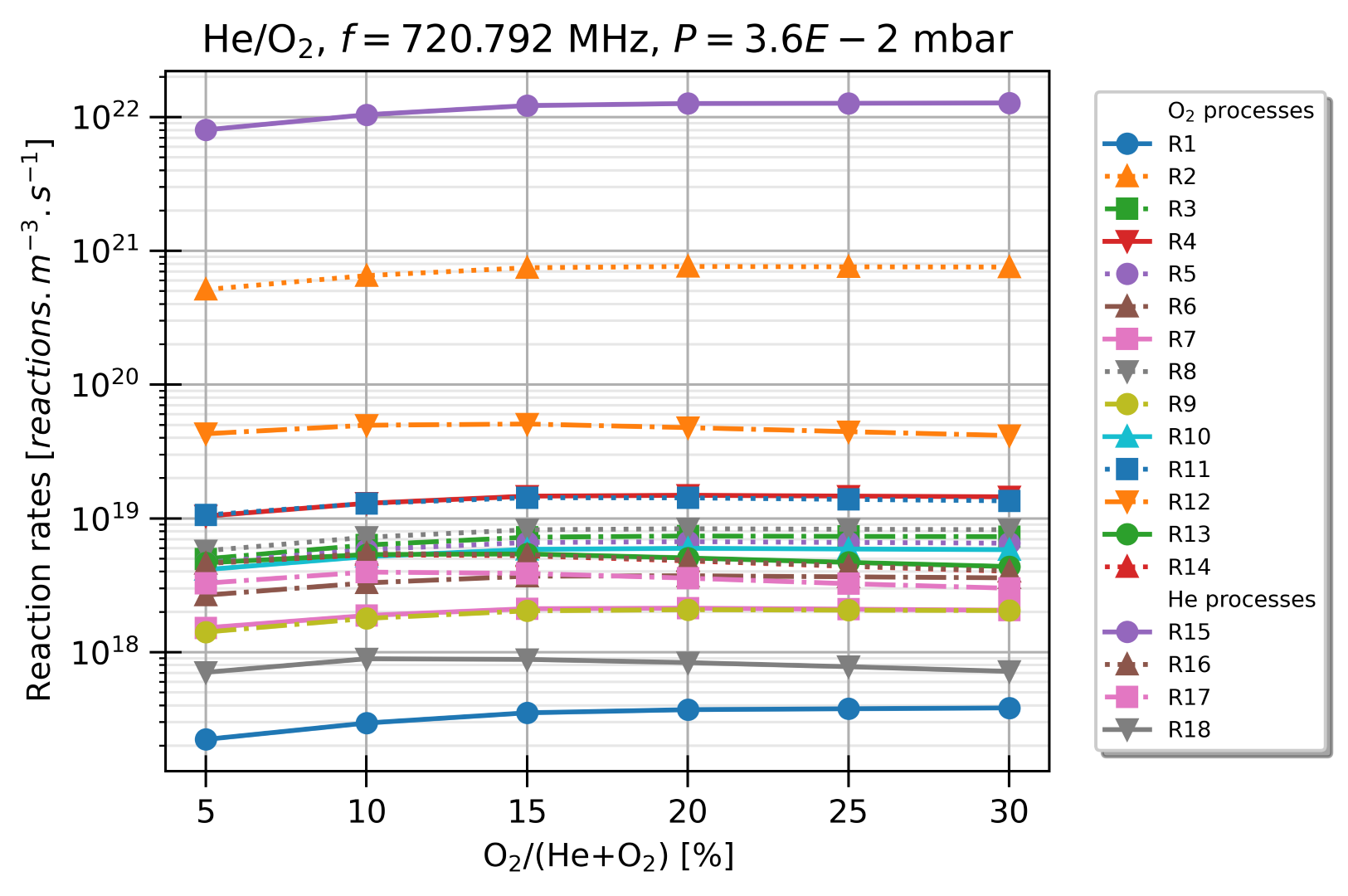}
         \caption{Electron impact reaction rates.}
         \label{fig17}
     \end{subfigure}
     \hfill
     \begin{subfigure}[t]{0.49\textwidth}
         \centering
         \includegraphics[width=\textwidth]{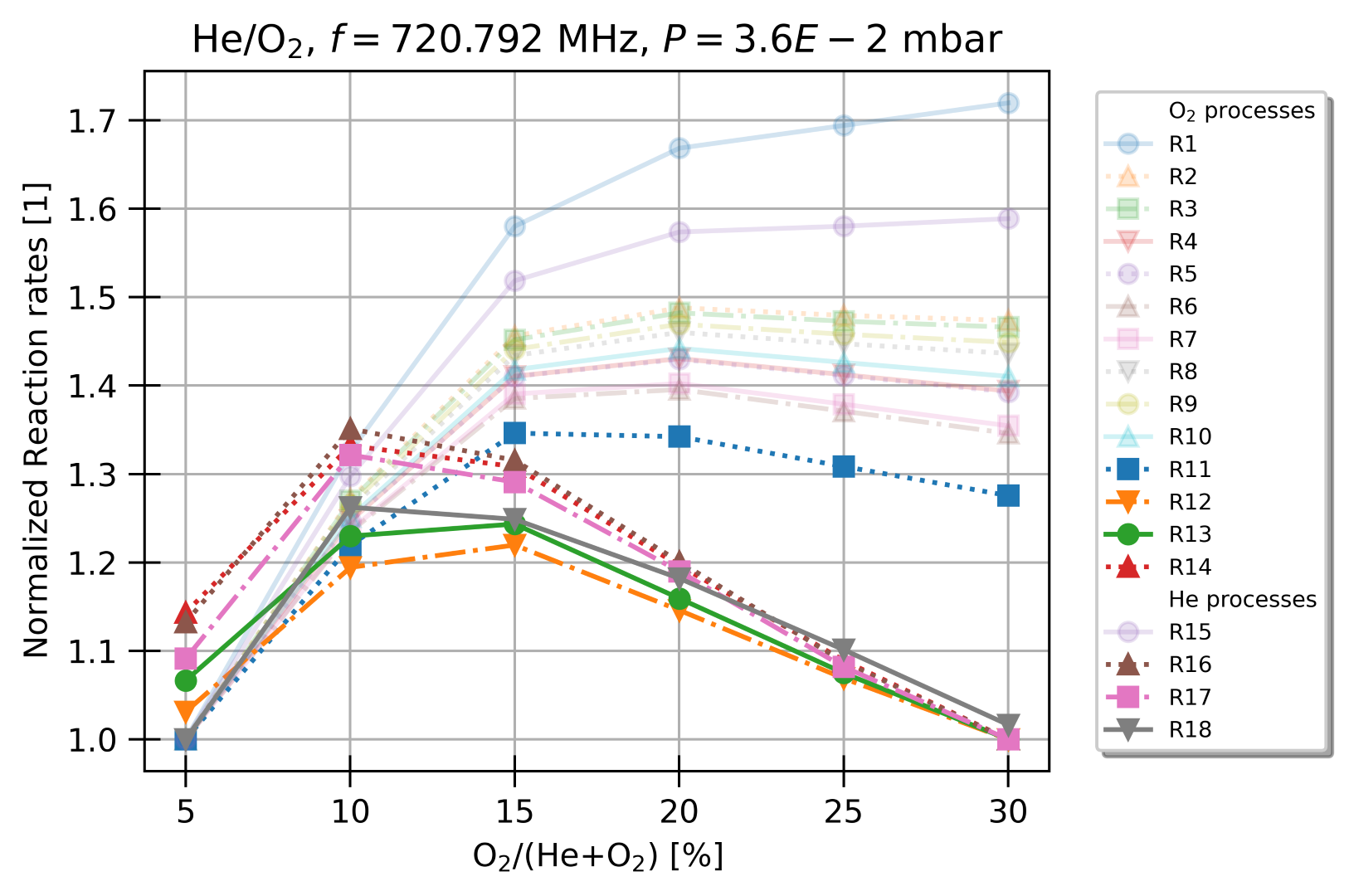}
	 \caption{Normalized electron impact reaction rates.}
         \label{fig18}
     \end{subfigure}
	 \caption{Electron impact reaction rates for O$_2$ and He processes computed from the EEDF and cross section data from the Biagi database for an He/O$_2$ plasma with varying oxygen concentration at constant pressure of $3.6\times10^{-2}$ mbar and constant frequency shift of $\Delta f=38$ MHz.}
\end{figure}

The resulting reaction rates are shown in Fig.~\ref{fig17}. Because the absolute values of the reaction rates span several orders of magnitude, normalized reaction rates are presented in Fig.~\ref{fig18} to facilitate comparison across the full range of oxygen concentrations. Each reaction rate is normalized to its minimum value over the investigated parameter space.



A comparison of Fig.~\ref{fig18} (highlighted reactions) with Fig.~\ref{fig15} reveals that both exhibit a bell-shaped dependence on the oxygen concentration, peaking close to 10-15\% of O$_2$ in He. This behavior is especially evident for reactions associated with the production of chemically reactive species, such as reactions R11, R12, R13 and R14, corresponding to O$_2$ excitation, dissociation and ionization processes. Reactions related to He such as R16, R17 and R18 follow a bell curve as well as high energy He excited states provides energy for dissociation and ionization of O$_2$.

Since the EEDF directly governs the plasma chemistry, this trend highlights the strong correlation between the measured cleaning rates and the computed electron-impact reaction rates. It should be noted, however, that the present analysis relies on a simplified kinetic model that includes only a limited set of electron-impact reactions. Contributions from heavy species processes, such as stepwise excitation, Penning ionization, or dissociation involving excited oxygen or helium states, are not included and may further influence the plasma chemistry under the studied conditions.

To summarize, the cleaning rate is governed by several coupled parameters: dilution gas, oxygen concentration, gas pressure, and frequency shift, each of which influences the others. Comparative Langmuir probe measurements between several gas mixtures, through the inferred EEDF, can be used to estimate the expected cleaning efficiency. In particular, Ar/O$_2$ plasmas were anticipated to be less effective than the other mixtures because of their depleted high-energy tail -- although showing similar electron densities -- and this expectation is confirmed by the cleaning rate data in Fig.~\ref{fig15}. 

The N$_2$/O$_2$ (25\%) mixture, which is close to air, yields the highest measured cleaning rate and thus appears to be the most effective mixture. However, it has not yet been employed, so its impact on cavity performance relative to other gases remains unquantified. 
The superior cleaning efficiency of N$_2$/O$_2$ can be attributed to the combined chemical reduction and oxidation effects of O$_2$ and N$_2$ on carbon-based contaminants \cite{app15137361}, along with a high electron density, a large population of high-energy electrons, and low pressure, which enhances the diffusion of reactive species to the walls.
Nonetheless, it has to be noted that nitrogen plasmas also create toxic substances such as cyanides (CN, HCN) and nitrous oxides (NO$_\text{x}$) when they react with organic contaminants \cite{app15137361}.

He/O$_2$ mixtures outperform Ar/O$_2$ mixtures -- despite operating at higher pressure -- in agreement with the improved cryogenic cavity performance reported by T.~Powers \textit{et al.} at Jefferson Lab \cite{powers:srf2023-wepwb054}. Taken together, these results indicate a higher cleaning potential for helium-based mixtures compared to argon-based ones, but they do not yet clarify whether the observed cavity performance gains arise solely from enhanced surface cleanliness or whether additional plasma-surface interaction mechanisms are involved. 



Hydrogen-based mixtures also show promise for efficient plasma cleaning of hydrocarbon contamination \cite{app15137361}, although they were not investigated in this study. Hydrogen is not only effective in removing hydrocarbon contamination but also chemically reduces niobium pentoxide (Nb$_2$O$_5$) to lower oxides (NbO$_2$ or NbO, and potentially even to metallic Nb) \cite{10.1116/1.569186}. Indeed, this process poses a risk by potentially contaminating the niobium lattice with hydrogen, as it removes the diffusion barrier provided by the native niobium pentoxide layer. Such hydrogen contamination could degrade superconducting performance, since niobium hydrides are known to deteriorate the performance of SRF cavities \cite{10.1063/1.1597367}.

\section{Conclusion \& Perspectives}
Simultaneous in-cavity measurements of the plasma parameters and cleaning rates were performed in this study. Despite the challenges imposed by the SRF cavity plasma reactor, a large parameter space was investigated, providing guidelines to improve the plasma processing effectiveness in SRF cavities. Numerical simulations combined with Langmuir probe measurements significantly increased the comprehension of the plasma mechanisms occurring in our atypical plasma reactor. The gas mixture plays an important role, but most importantly, low pressure (i.e. enhanced diffusion) and high electron density (i.e. large frequency shift) proved to yield the highest cleaning rates -- independently of the gas mixture. Still, pressure dependent studies of optimum O$_2$ content for a set of gases lack to this study, but this would represent a huge experimental effort.

Cavity performance improvement subsequent to plasma processing is up to now attributed to improved surface cleanliness. In the case of niobium cavities, plasma processing can potentially further improve superconducting performance if plasma-surface interactions are better understood. Surface analysis of niobium samples exposed to plasma is probably the next step to unlock further performance gain. This work has been initiated in \cite{giaccone:srf2021-wepcav001, ZHANG2019143} but with a vacuum break between plasma exposure and surface analysis. Furthermore, with the present methodology, plasma processing targets only hydrocarbon-based contamination. To remove other contaminants such as particulates, it would require sputtering from plasma ions, provided that the plasma potential can be controlled and increased to allow ions to gain more energy in the sheath potential drop.



\ack{The authors acknowledge IJCLab's staff at the Vacuum \& Surfaces platform for the technical support. All the measurements presented in this study were performed at the Vacuum \& Surfaces platform. For more information on the Vacuum and Surfaces platform, see \url{https://www.ijclab.in2p3.fr/en/platforms/vacuum-and-surfaces-platform/}.

We thank the CNRS Cold Plasmas Network for providing the ``Quë-Do'' Langmuir probe electronics and expertise. We also thank the Laboratoire de Physique des Plasmas (LPP) for providing the Impedans ALP Langmuir probe electronics.

We thank the plasma team at Jefferson Lab for the training on their plasma processing test stand and for the training on COMSOL plasma simulation, especially N. Raut.

Finally, we thank the plasma teams at Argone, Brookhaven, Fermilab, FRIB and JLab for the useful discussions, information sharing and suggestions.}

\funding{This work benefited from funding from IN2P3 and the French State aid under France 2030 (P2I - Graduate School Physique) with the reference ANR-11-IDEX-0003. This work was partially supported by the European Union’s Horizon Europe Marie Sklodowska-Curie Staff Exchanges programme under grant agreement no. 101086276.}

\roles{\textbf{C. CHENEY:} Writing -- original draft, Investigation, Data Curation, Formal analysis, Visualization, Methodology, Software, Conceptualization, Resources.
\textbf{G. ABI-ABBOUD:} Investigation, Data Curation, Formal Analysis.
\textbf{S. BÉCHU:} Writing -- review \& editing, Resources, Software, Methodology, Validation.
\textbf{A. BÈS:} Resources, Software.
\textbf{L. BONNY:} Resources, Software.
\textbf{T. Gerardin:} Resources.
\textbf{B. MERCIER:} Resources.
\textbf{E. MISTRETTA:} Resources.
\textbf{J. YEMANE:} Resources.
\textbf{D. LONGUEVERGNE:} Writing -- review \& editing, Validation, Supervision, Resources, Project administration, Methodology, Funding acquisition, Conceptualization.
}


\data{The data that support the findings of this study are available from the corresponding author, C. CHENEY, upon reasonable request.}


\appendix

\section{Axial distribution of the plasma parameters \& electron kinetic}\label{appendixA}
This appendix describes the spatial distribution of plasma parameters along the beam axis, combining experimental measurements and numerical simulations. It also provides an analysis of electron kinetics in the uncommon SRF cavity plasma reactor.

Despite the plasma instabilities discussed in section \ref{subsect4.1.1}, a spatial study of plasma parameters was conducted for a pure argon plasma confined to a single accelerating gap. This configuration was chosen to avoid instabilities during probe insertion into the plasma core, with measurements performed at a relatively large frequency shift. The results of both probe measurements and numerical simulations are presented below.

The probe was moved along the beam tube axis over a 4 cm range. The position $x=0$ corresponds to the accelerating gap center -- in other words, the plasma discharge center. The position $x=-L/2=-6.55/2=-3.275$ [cm] is referenced as the entry in the accelerating gap. $L=6.55$ cm being the accelerating gap length. However, excessive insertion of the probe into the accelerating gap disturbed the plasma, rendering measurements non-representative due to the ignition of a secondary plasma near the probe tip. This phenomenon is attributed to the enhancement of the electric field around the probe tip when inserted too deeply, as the electric field norm peaks at the center of the accelerating gap.

\begin{figure}[htb!]
\centering
     \begin{subfigure}[t]{0.35\textwidth}
         \centering
         \includegraphics[width=\textwidth]{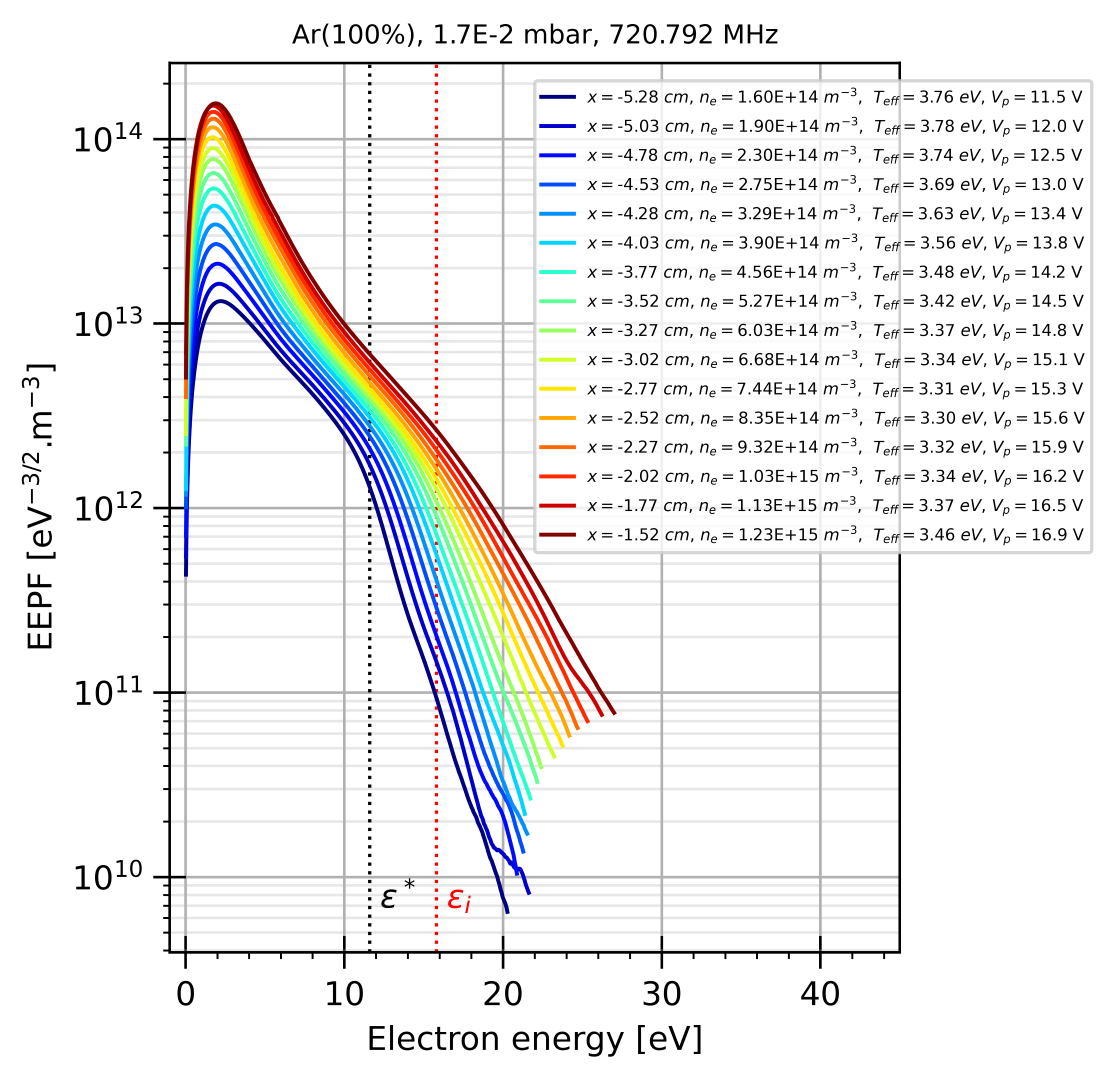}
         \caption{EEPF}
         \label{fig4a}
     \end{subfigure}
     \hfill
     \begin{subfigure}[t]{0.3\textwidth}
         \centering
         \includegraphics[width=\textwidth]{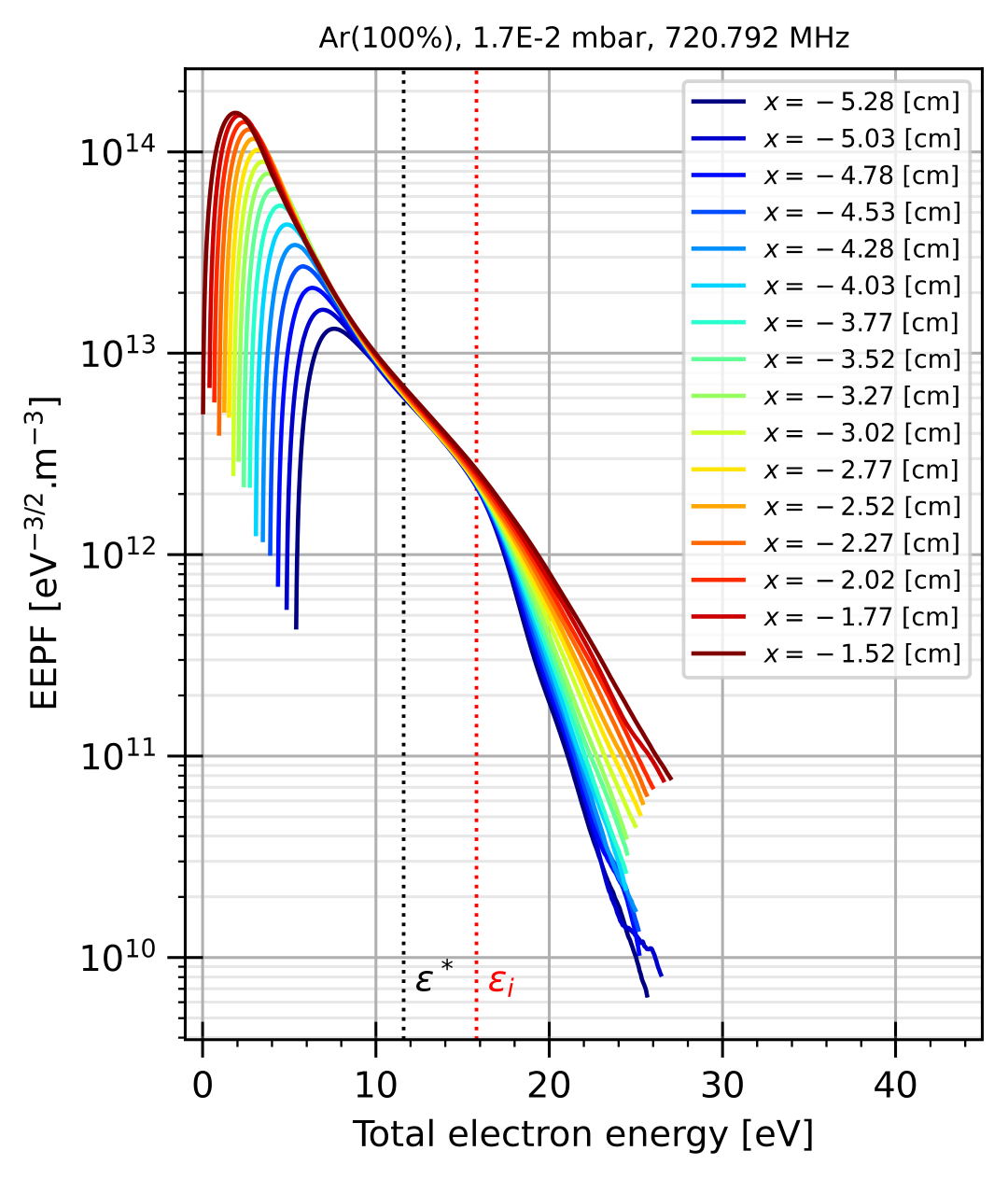}
         \caption{EEPF, total electron energy}
         \label{fig4b}
     \end{subfigure}
     \hfill
     \begin{subfigure}[t]{0.3\textwidth}
         \centering
         \includegraphics[width=\textwidth]{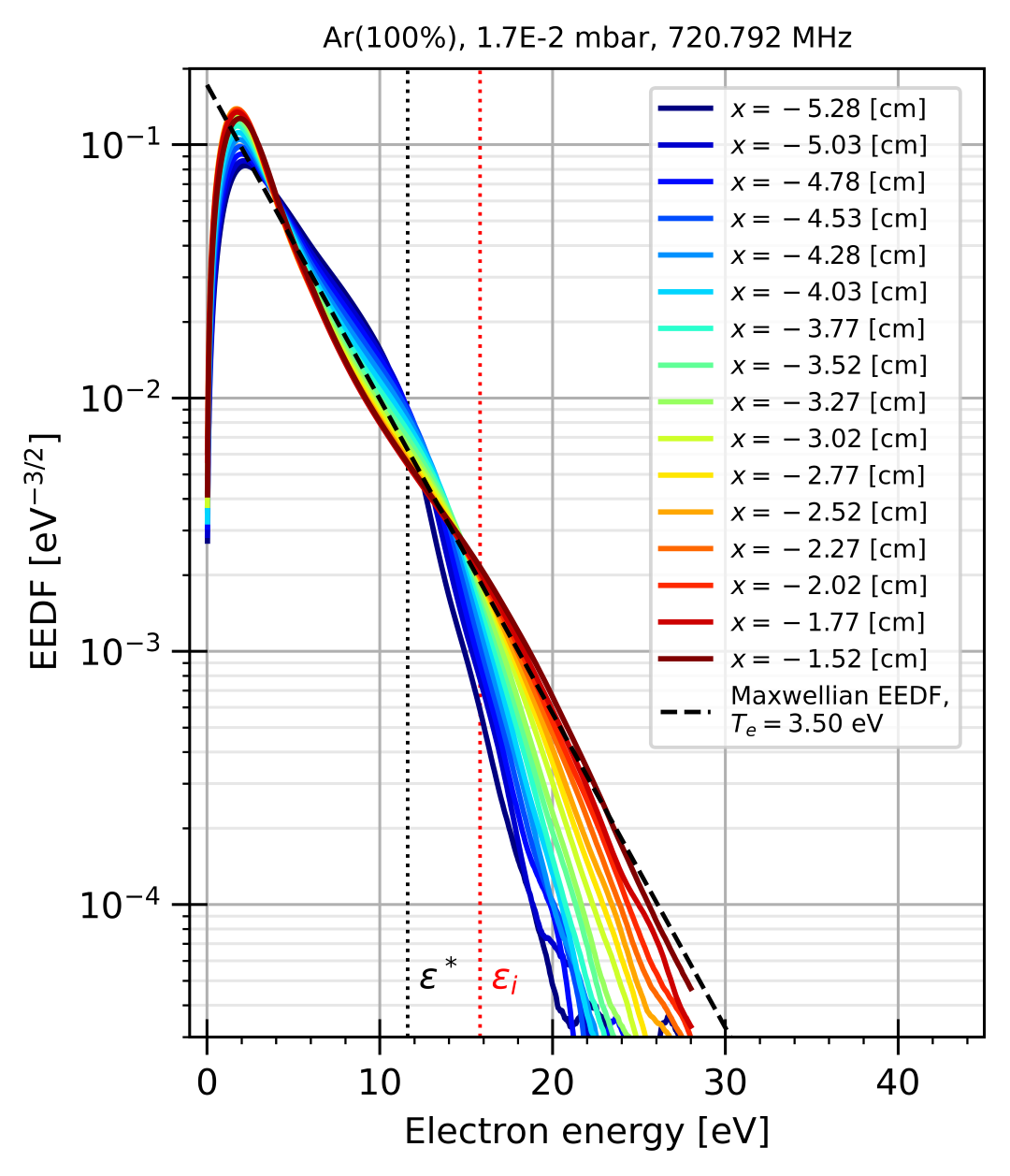}
         \caption{EEDF}
         \label{fig4c}
     \end{subfigure}
	 \caption{EEPF and EEDF evolution as a function of the $x$ position along the cavity beam axis for a pure Ar plasma at a pressure of $1.7\times10^{-2}$ mbar and a frequency shift $\Delta f=38$ MHz.}
	 \label{fig4}
\end{figure}

EEPFs and EEDFs for a pure argon plasma at $1.7\times10^{-2}$ mbar and a frequency shift of 38 MHz at different probe locations are represented in Fig. \ref{fig4}. At the discharge edge, where the probe is fully out ($x=-5.28$ cm), a bi-Maxwellian distribution is observed: a first group of low energy electrons (below 5 eV) is characterized by a low electron temperature, and a second group of high energy electron (over 5 eV) population with a high temperature. When the probe is further inserted in the discharge, the EEDF evolves toward a three-temperature structure with a well-expressed low energy bump and a fall in the inelastic energy range. 

\begin{figure}[htb!]
\centering
     \begin{subfigure}[t]{0.45\textwidth}
         \centering
         \includegraphics[width=\textwidth]{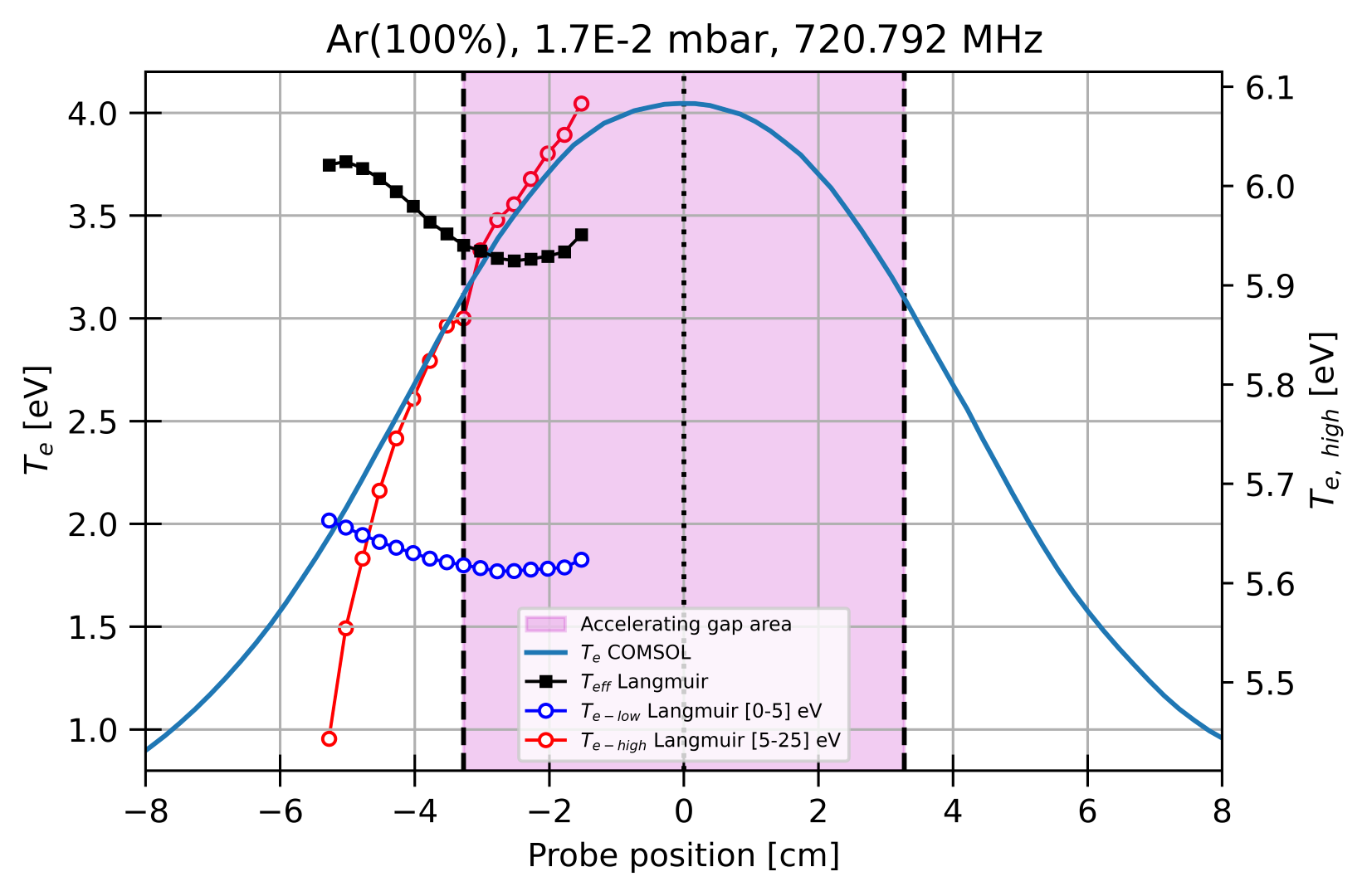}
         \caption{Measured effective electron temperature $T_{eff}$, low energy range temperature $T_{e-low}$ and high energy range temperature $T_{e-high}$ (secondary y-axis) at different axial location and COMSOL simulation data for comparison.}
         \label{fig5a}
     \end{subfigure}
     \hfill
     \begin{subfigure}[t]{0.45\textwidth}
         \centering
         \includegraphics[width=\textwidth]{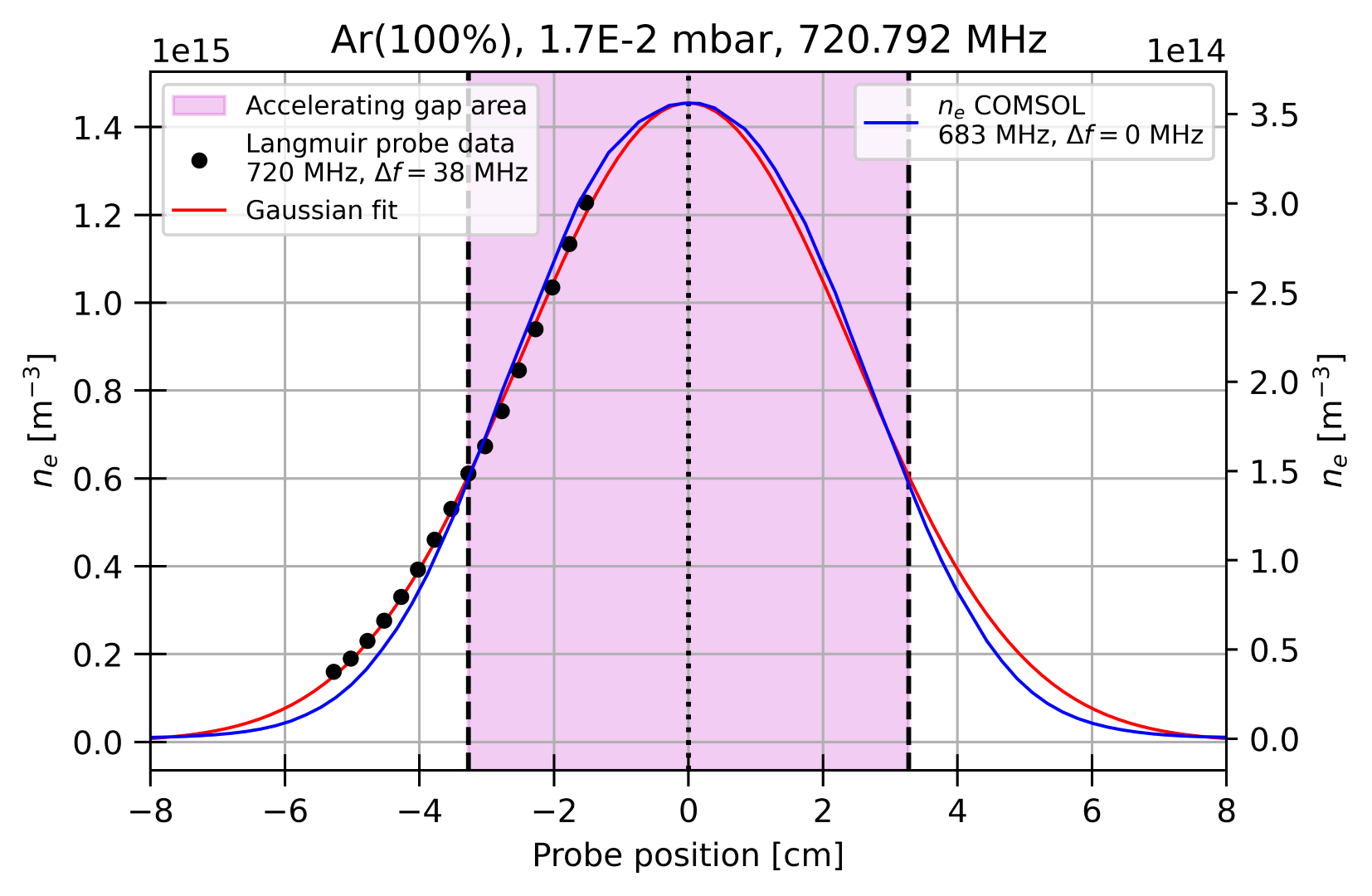}
         \caption{Measured electron density ($n_e$) at different axial location and COMSOL simulation data for comparison on the secondary y-axis.}
         \label{fig5b}
     \end{subfigure}
	 \caption{$T_e$ and $n_e$ evolution as a function of the $x$ position along the cavity beam axis for a pure Ar plasma at a pressure of $1.7\times10^{-2}$ mbar and a frequency shift $\Delta f=38$ MHz. COMSOL simulation data comes from a 3D simulation with pure Ar at $1.5\times10^{-2}$ mbar, a frequency shift $\Delta f=0$ MHz and at time $t=10$ seconds, when the plasma achieve equilibrium.
	 }
	 \label{fig5}
\end{figure}

The temperatures for the low and high energy range are explicitly shown in Fig. \ref{fig5a}. There, the effective electron temperature $T_{e}$ (or $T_{eff}$) is determined through integration of the EEDF based on the Druyvesteyn method, following equation (\ref{equation2}). $T_{e-low}$ and $T_{e-high}$ are determined through fitting with multiple Gaussian on the [0--5] eV, [5--25] eV range respectively.

The effective electron temperature spatial profile exhibits a concave shape in Fig. \ref{fig5a} with a temperature slightly higher at the discharge edge than at the discharge center. This can be explained by the confinement of low energy electrons at the discharge center due to the ambipolar potential. Indeed, low energy electrons overpopulate the EEDF and create a bump in the low energy range as shown in Fig. \ref{fig4a}. This bump artificially cools the effective electron temperature at the discharge center. This is why the measured $T_e$ does not follow the numerical model trend, but $T_{e-high}$ do, with a temperature increasing toward the discharge center. Also, the fluid model with an assumed EEDF cannot capture this kinetic effect.

In Fig.~\ref{fig4a}, the high-energy tail of the EEPF is depleted. The change in the EEPF slope occurs near the local plasma potential, suggesting that the primary electron energy loss mechanism is the loss of high-energy electrons to the walls \cite{Godyak_eedf_icp}. Only electrons with sufficient energy to overcome the plasma potential barrier are lost to the walls.

Fig. \ref{fig4b} represents the EEPFs in the total electron energy scale (kinetic plus potential energy) at various axial positions. It can be seen that the EEPF shape is the same in the [8-16] eV total energy interval, but the high-energy tail drops significantly with the axial distance from the discharge center. This suggests that electrons are mainly heated near the discharge center, where the electric field is the highest.



As the plasma potential decreases from the discharge center toward the wall, the potential energy of electrons -- referenced to the center where it is set to zero -- increases as they move outward. Consequently, the kinetic energy scale of the measured EEPFs must be shifted by the local plasma potential variation, as illustrated in Fig. \ref{fig4b}. This shift reflects the increase in potential energy acquired by electrons traveling away from the discharge center. As a result, the total energy corresponding to the low-energy cutoff of the EEPFs increases from the center toward the discharge edge. The decreasing plasma potential along the beam axis creates an ambipolar potential barrier that low-energy electrons generated by ionization cannot overcome, leading to a truncation of the EEPFs in the low-energy range.

 

Similar EEDFs to those in Fig.~\ref{fig4a} and \ref{fig4b} are discussed in \cite{10.1063/1.4989704, 10.1063/1.3692768, 10.1063/1.4986495}. These studies present spatial measurements of plasma parameters and EEPFs in inductively coupled plasmas (ICPs) along the axial axis. They observed that EEPFs coincide in the total electron energy scale over the entire energy range, concluding that this indicates a non-local electron kinetic regime. The EEPFs in Fig.~\ref{fig4} closely resemble those in the literature, except that in the SRF cavity reactor, the high-energy tail does not coincide at all positions (Fig.~\ref{fig4b}). This discrepancy is believed to result from the electric field gradient along the measurement axis. Since the electron density is below the critical density, the radio frequency field can penetrate the plasma bulk without attenuation. Electrons are thus more heated at the discharge center, where the electric field is highest, explaining why the temperature of the high-energy tail decreases toward the discharge edge. In other words, the spatial evolution of EEPFs exhibits both non-local and local electron kinetics: non-local for the low-energy group ($\varepsilon_{\text{tot}} < \varepsilon_i$) and local for the high-energy group ($\varepsilon_{\text{tot}} > \varepsilon_i$). Such dual behavior was observed by Godyak and Piejak in a capacitively coupled plasma (CCP) discharge \cite{10.1063/1.110227} and may be explained by a significant difference in the electron energy relaxation length ($\lambda_\varepsilon$) for the low and high-energy electron groups.

\begin{figure}[htb!]
\centering
     \begin{subfigure}[t]{0.25\textwidth}
         \centering
         \includegraphics[width=\textwidth]{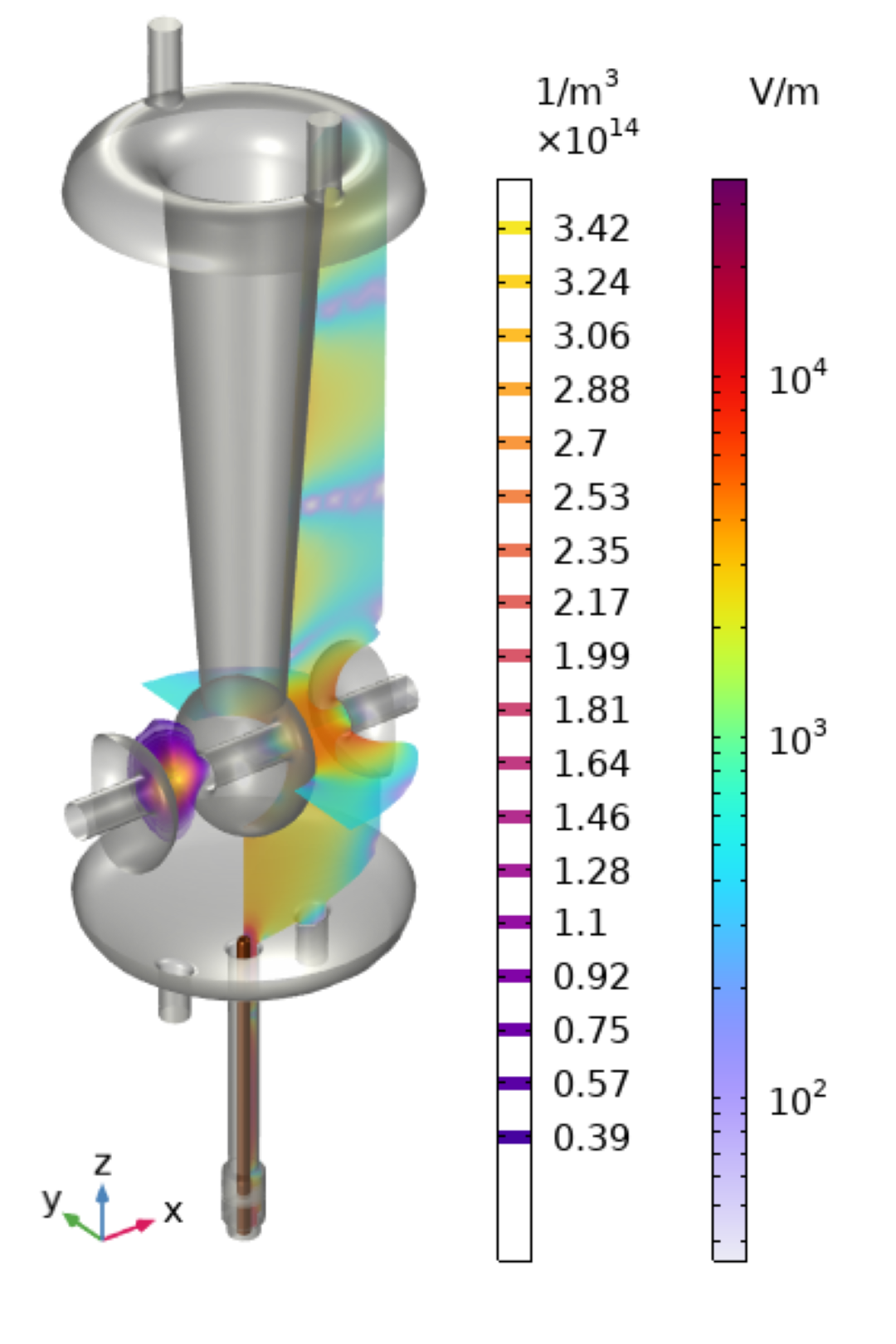}
         \caption{3D representation of the electron density and electric field norm.}
         \label{fig6a}
     \end{subfigure}
     \hfill
     \begin{subfigure}[t]{0.73\textwidth}
         \centering
         \includegraphics[width=\textwidth]{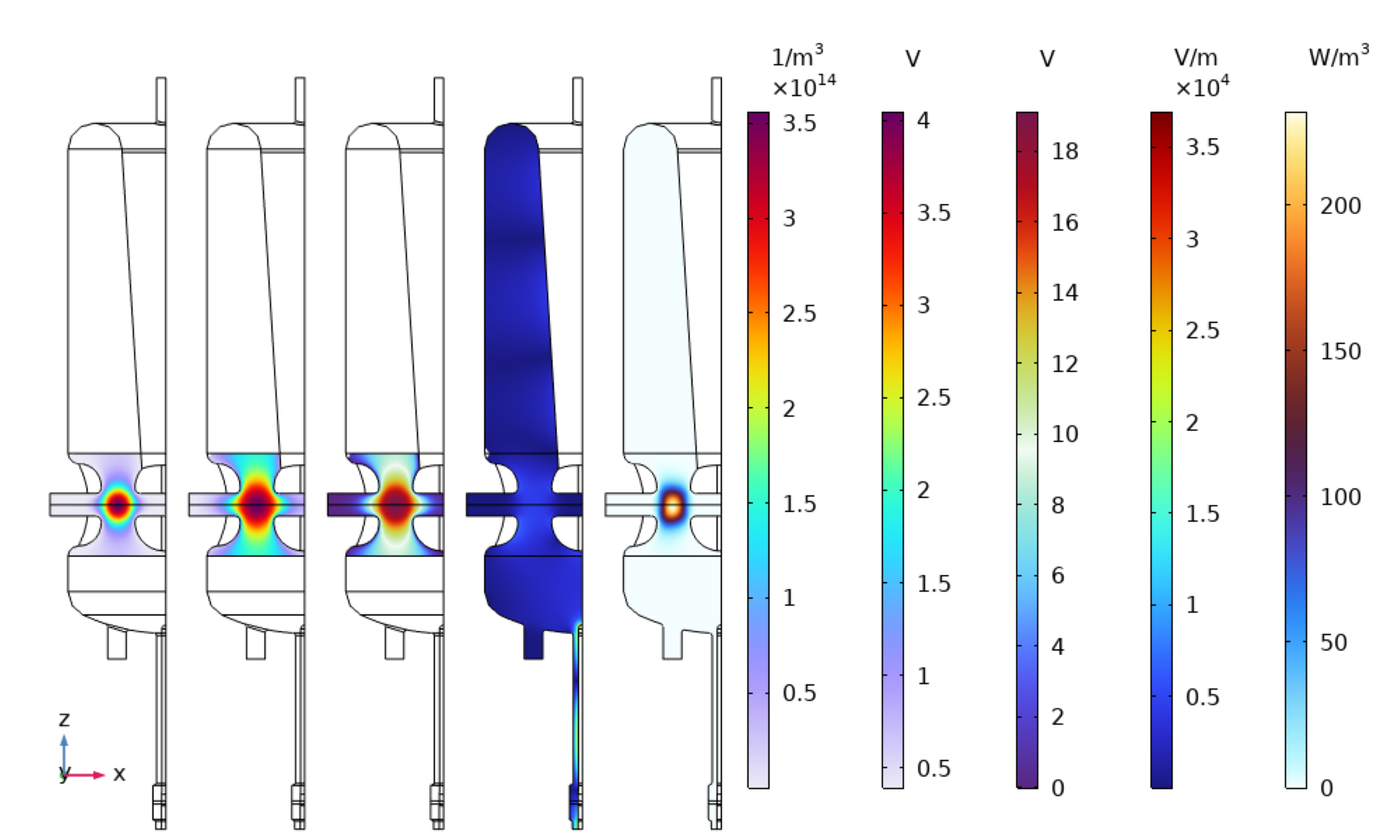}
         \caption{From left to right with corresponding legend: electron density $n_e$, electron temperature $T_e$, plasma potential $V_p$, electric field norm $|E|$, power deposition in plasma (resistive losses).}
         \label{fig6b}
     \end{subfigure}
	 \caption{Numerical simulation results for an argon plasma at $1.5\times10^{-2}$ mbar and a frequency shift $\Delta f=0$ MHz. The RF power is set to 80 W.}
	 \label{fig6}
\end{figure}

Fig.~\ref{fig5b} shows the electron density profile along the beam axis. The measured $n_e$ fits well with a Gaussian distribution and aligns closely with numerical simulation results. While the absolute values of $n_e$ differ between experiments and simulations due to differing frequency shifts ($\Delta f = 38$~MHz for measurements and $\Delta f = 0$~MHz for simulations), the profiles remain similar and are expected to retain this Gaussian-like shape throughout the frequency shift process. The numerical model is detailed in the following section. 3D and 2D maps of the simulation results are presented in Fig.~\ref{fig6}.


\paragraph{Numerical model} 
To compare Langmuir probe measurements, a numerical model was developed using the COMSOL Multiphysics plasma module. This self-consistent fluid model is coupled with Maxwell's equations and is based on the following COMSOL tutorials \cite{COMSOL_MW_plasma, COMSOL_MW_cavity_plasma}. Due to the computational challenges of 3D simulations, only a quarter of the cavity was simulated, leveraging its symmetrical geometry. Additionally, the plasma domain was restricted to a small portion of the cavity volume to further reduce computation time.

In the model, the EEDF is assumed to follow a generalized form with $g = 1.3$ (Eq.~\ref{equation6}). This assumed EEDF aligns more closely with the measured EEDFs, as it accounts for the depleted high-energy tail compared to a Maxwellian EEDF (Fig.~\ref{fig7b}). When a Maxwellian EEDF was assumed in the model, the results did not match experimental observations, as the plasma failed to ignite in the accelerating gap.

\begin{equation}
f(\varepsilon)=\varphi^{-3/2} \beta_1 exp \left(-\left(\frac{\varepsilon \beta_2}{\varphi}\right)^g\right)
\label{equation6}
\end{equation}

\[
\beta_1=\Gamma(5/2g)^{3/2}\Gamma(3/2g)^{-5/2},~~\beta_2=\Gamma(5/2g)\Gamma(3/2g)^{-1}
\]

Where, $\varepsilon$ is the electron energy, (eV);  $\varphi$ is the mean electron energy, (eV); and $g$ is a factor between 1 and 2. $g=1$ corresponds to a Maxwellian EEDF and $g=2$ to a Druyvesteyn EEDF.

In Fig.~\ref{fig6b}, the plasma parameters are shown in the plane perpendicular to the beam axis. A key parameter is the power deposition, or resistive heating, which represents the power transferred from the electromagnetic fields to the electrons. The plot confirms that power is primarily deposited at the discharge center, consistent with a "wave-heated discharge" where coupling is electromagnetic. This observation highlights the significant role of the spatial electric field distribution in controlling the plasma's spatial distribution.

\printbibliography

@misc{martinello2022plasmaprocessinginsitufield,
      title={Plasma Processing for In-Situ Field Emission Mitigation of Superconducting Radiofrequency (SRF) Cryomodules}, 
      author={M. Martinello and P. Berrutti and B. Giaccone and S. Belomestnykh and M. Checchin and G. V. Eremeev and A. Grassellino and T. Khabibouilline and A. Netepenko and R. Pilipenko and A. Romanenko and S. Posen and G. Wu and D. Gonnella and M. Ross and J. T. Maniscalco and T. Powers},
      year={2022},
      eprint={2203.12442},
      archivePrefix={arXiv},
      primaryClass={physics.acc-ph},
      url={https://arxiv.org/abs/2203.12442}, 
}

@article{TYAGI201629,
title = {Improving the work function of the niobium surface of SRF cavities by plasma processing},
journal = {Applied Surface Science},
volume = {369},
pages = {29-35},
year = {2016},
issn = {0169-4332},
doi = {https://doi.org/10.1016/j.apsusc.2016.02.030},
url = {https://www.sciencedirect.com/science/article/pii/S0169433216301908},
author = {P.V. Tyagi and M. Doleans and B. Hannah and R. Afanador and C. McMahan and S. Stewart and J. Mammosser and M. Howell and J. Saunders and B. Degraff and S.-H. Kim}
}

@inproceedings{Doleans2013PlasmaPR,
  title={Plasma Processing R\&D for the SNS Superconducting Linac RF Cavities},
  author={Marc Doleans and R. Afanador and Jeffrey Allen Ball and Willem Blokland and Mark Crofford and Brian Degraff and David Douglas and Brian Hannah and Matthew Howell and Christopher McMahan and Jeffrey Saunders and Puneet Tyagi},
  year={2013},
  url={https://api.semanticscholar.org/CorpusID:51805469}
 }

@inproceedings{Tyagi:2014bwa,
    author = "Tyagi, Puneet and Doleans, Marc and Kim, Sang-Ho and Afanador, Ralph and McMahan, Christopher",
    title = "{Plasma Processing of Nb Surfaces for SRF Cavities}",
    booktitle = "{27th International Linear Accelerator Conference}",
    pages = "MOPP115",
    year = "2014"
}

@article{10.1063/1.4972838,
    author = {Doleans, Marc},
    title = {Ignition and monitoring technique for plasma processing of multicell superconducting radio-frequency cavities},
    journal = {Journal of Applied Physics},
    volume = {120},
    number = {24},
    pages = {243301},
    year = {2016},
    month = {12},
    issn = {0021-8979},
    doi = {10.1063/1.4972838},
    url = {https://doi.org/10.1063/1.4972838},
    eprint = {https://pubs.aip.org/aip/jap/article-pdf/doi/10.1063/1.4972838/15186175/243301_1_online.pdf},
}

@online{Plaquette_Que_Do,
  author = {plasmas-froids.cnrs.fr},
  title = {Dispositif de mesure des paramètres plasma par sonde de Langmuir: QUË-DO},
  year = 2020,
  url = {https://plasmas-froids.cnrs.fr/wp-content/uploads/2020/12/Plaquette_Sonde.pdf},
  urldate = {2025-10-30}
}

@article{druyvesteyn1930niedervoltbogen,
  title={Der niedervoltbogen},
  author={Druyvesteyn, Mari Johan},
  journal={Zeitschrift f{\"u}r Physik},
  volume={64},
  number={11},
  pages={781--798},
  year={1930},
  publisher={Springer}
 }

@misc{BechuEtudeSonde,
author = {S. B{\'e}chu and A. B{\`e}s and L. Bonny and D. Fombaron and A. Lacoste},
  title = {Étude de plasmas par sondes électrostatiques},
  url = {https://plasmas-froids.cnrs.fr/wp-content/uploads/2021/01/Sondes_Electrostatiques_Bechu_2.pdf},
  urldate = {2025-10-30}
}

@article{K_Popov_2006,
doi = {10.1088/1742-6596/44/1/007},
url = {https://doi.org/10.1088/1742-6596/44/1/007},
year = {2006},
volume = {44},
number = {1},
pages = {60},
author = {Tsv K Popov and M Dimitrova and F M Dias and V N Tsaneva and N A Stelmashenko and M G Blamire and Z H Barber},
title = {Second derivative Langmuir probe diagnostics of gas discharge plasma at intermediate pressures (review article)},
journal = {Journal of Physics: Conference Series}
}

@software{comsol,
  author = {{COMSOL Multiphysics®}},
  title = {COMSOL AB, Stockholm, Sweden.},
  url = {www.comsol.com.},
  version = {6.2},
}

@inproceedings{hartung:srf2023-thixa01,
  author       = {W. Hartung and W. Chang and K. Elliott and S.H. Kim and T. Konomi and K. Saito and P.R. Tutt and T. Xu},
% author       = {W. Hartung and W. Chang and K. Elliott and S.H. Kim and T. Konomi and K. Saito and others},
% author       = {W. Hartung and others},
  title        = {{Investigation of Plasma Processing for Coaxial Resonators}},
% booktitle    = {Proc. SRF'23},
  booktitle    = {Proc. 21th Int. Conf. RF Supercond. (SRF'23)},
  pages        = {960--967},
  eid          = {THIXA01},
  language     = {english},
  keywords     = {plasma, cavity, SRF, HOM, coupling},
  venue        = {Grand Rapids, MI, USA},
  series       = {International Conference on RF Superconductivity},
  number       = {21},
  publisher    = {JACoW Publishing, Geneva, Switzerland},
  month        = {09},
  year         = {2023},
  issn         = {2673-5504},
  isbn         = {978-3-95450-234-9},
  doi          = {10.18429/JACoW-SRF2023-THIXA01},
  url          = {https://jacow.org/srf2023/papers/thixa01.pdf},
  abstract     = {{Plasma processing has been investigated by several facilities as a method to mitigate degradation of SRF cavity performance. It provides an alternative to removal and disassembly of cryomodules for refurbishment of each cavity via repeat etching and rinsing. Promising results have been obtained by several groups. Studies of plasma processing for quarter-wave resonators (QWRs) and half-wave resonators (HWRs) were undertaken at FRIB, where a total of 324 such resonators are presently in operation. Plasma ignition and optimization measurements were done with room-temperature-matched input couplers. Plasma cleaning tests were done on several QWRs using the fundamental power coupler (FPC) to drive the plasma. We investigated the usefulness of higher-order modes (HOMs) to drive the plasma. HOMs allow for less mismatch at the FPC and hence lower field in the coupler relative to the cavity. Before-and-after cold tests showed a significant reduction in field emission X-rays with judicious application of plasma processing.}},
}

@InProceedings{wu:srf2019-mop085,
  author       = {A.D. Wu and Q.W. Chu and H. Guo and Y. He and S.C. Huang and T.C. Jiang and C.L. Li and Z.Q. Lin and F. Pan and Y.K. Song and T. Tan and W.M. Yue and S.H. Zhang and H.W. Zhao},
% author       = {A.D. Wu and Q.W. Chu and H. Guo and Y. He and S.C. Huang and T.C. Jiang and others},
% author       = {A.D. Wu and others},
  title        = {{The Destructive Effects to the RF Coupler by the Plasma Discharge}},
  booktitle    = {Proc. SRF'19},
  pages        = {285--287},
  paper        = {MOP085},
  language     = {english},
  keywords     = {plasma, cavity, experiment, vacuum, coupling},
  venue        = {Dresden, Germany},
  series       = {International Conference on RF Superconductivity},
  number       = {19},
  publisher    = {JACoW Publishing, Geneva, Switzerland},
  month        = {aug},
  year         = {2019},
  issn         = {""},
  isbn         = {978-3-95450-211-0},
  doi          = {10.18429/JACoW-SRF2019-MOP085},
  url          = {http://jacow.org/srf2019/papers/mop085.pdf},
  note         = {https://doi.org/10.18429/JACoW-SRF2019-MOP085},
}

@inproceedings{cheney,
    author  = {C. Cheney and al},
    title   = "Overview of Plasma Processing Activities at IJCLab",
    publisher = "presented at \emph{TTC2023 Meeting}, Fermilab, Dec. 2023",
    url = {https://indico.fnal.gov/event/60446/contributions/277987/attachments/173134/248424/TTC_meeting_plasma_IJCLab_CHENEY_updated.pdf}
}

@inproceedings{hartung:hiat2025-tuc03,
    author = {W. Hartung and others.},
    title = {Development of plasma processing for superconducting half-wave resonators},
    booktitle = {Proc. HIAT2025},
    %  booktitle = {Proc. 16th International Conference on Heavy Ion Accelerator Technology},
    pages = {118-121},
    paper = {TUC03},
    venue = { East Lansing, MI, USA},
    series = {HIAT},
    number = {16},
    publisher = {JACoW Publishing, Geneva, Switzerland},
    month = {06},
    year = {2025},
    issn = {2673-5547},
    isbn = {978-3-95450-260-8},
    doi = {10.18429/JACoW-HIAT2025-TUC03},
    url = {https://indico.jacow.org/event/82/contributions/8940},
    language = {English},
    %eventdate = {2025-06-22/2025-06-27}
}

@article{Godyak_eedf_icp,
author = {Godyak, V. and Piejak, R and Alexandrovich, Benjamin},
year = {2002},
month = {11},
pages = {525},
title = {Electron energy distribution function measurements and plasma parameters in inductively coupled argon plasma},
volume = {11},
journal = {Plasma Sources Science and Technology},
doi = {10.1088/0963-0252/11/4/320}
}

@article{10.1063/1.4989704,
    author = {Gao, Fei and Zhang, Yu-Ru and Li, Hong and Liu, Yang and Wang, You-Nian},
    title = {Spatial distributions of plasma parameters in inductively coupled hydrogen discharges with an expansion region},
    journal = {Physics of Plasmas},
    volume = {24},
    number = {7},
    pages = {073508},
    year = {2017},
    month = {07},
    issn = {1070-664X},
    doi = {10.1063/1.4989704},
    url = {https://doi.org/10.1063/1.4989704},
    eprint = {https://pubs.aip.org/aip/pop/article-pdf/doi/10.1063/1.4989704/16003557/073508_1_online.pdf},
}

@article{10.1063/1.3692768,
    author = {Lee, Hyo-Chang and Chung, Chin-Wook},
    title = {Experimental measurements of spatial plasma potentials and electron energy distributions in inductively coupled plasma under weakly collisional and nonlocal electron kinetic regimes},
    journal = {Physics of Plasmas},
    volume = {19},
    number = {3},
    pages = {033514},
    year = {2012},
    month = {03},
    issn = {1070-664X},
    doi = {10.1063/1.3692768},
    url = {https://doi.org/10.1063/1.3692768},
    eprint = {https://pubs.aip.org/aip/pop/article-pdf/doi/10.1063/1.3692768/15652998/033514_1_online.pdf},
}

@article{10.1063/1.4986495,
    author = {Li, Hong and Liu, Yang and Zhang, Yu-Ru and Gao, Fei and Wang, You-Nian},
    title = {Nonlocal electron kinetics and spatial transport in radio-frequency two-chamber inductively coupled plasmas with argon discharges},
    journal = {Journal of Applied Physics},
    volume = {121},
    number = {23},
    pages = {233302},
    year = {2017},
    month = {06},
    issn = {0021-8979},
    doi = {10.1063/1.4986495},
    url = {https://doi.org/10.1063/1.4986495},
    eprint = {https://pubs.aip.org/aip/jap/article-pdf/doi/10.1063/1.4986495/13180014/233302_1_online.pdf},
}

@article{COMSOL_MW_plasma,
    author = {COMSOL Inc},
    journal = {COMSOL Application Gallery},
    title = {In-Plane Microwave Plasma},
    url = {https://www.comsol.com/model/in-plane-microwave-plasma-8664},
}

@article{COMSOL_MW_cavity_plasma,
    author = {COMSOL Inc},
    journal = {COMSOL Application Gallery},
    title = {Microwave Cavity Plasma Reactor},
    url = {https://www.comsol.com/model/microwave-cavity-plasma-reactor-115681},

}

@article{10.1063/1.110227,
    author = {Godyak, V. A. and Piejak, R. B.},
    title = {Paradoxical spatial distribution of the electron temperature in a low pressure rf discharge},
    journal = {Applied Physics Letters},
    volume = {63},
    number = {23},
    pages = {3137-3139},
    year = {1993},
    month = {12},
    issn = {0003-6951},
    doi = {10.1063/1.110227},
    url = {https://doi.org/10.1063/1.110227},
    eprint = {https://pubs.aip.org/aip/apl/article-pdf/63/23/3137/18499750/3137_1_online.pdf},
}

@inproceedings{powers:srf2023-wepwb054,
  author       = {T. Powers and N.C. Brock and T.D. Ganey},
  title        = {{In Situ Plasma Processing of Superconducting Cavities at JLab, 2023 Update}},
% booktitle    = {Proc. SRF'23},
  booktitle    = {Proc. 21th Int. Conf. RF Supercond. (SRF'23)},
  pages        = {701--705},
  eid          = {WEPWB054},
  language     = {english},
  keywords     = {cavity, plasma, cryomodule, HOM, radiation},
  venue        = {Grand Rapids, MI, USA},
  series       = {International Conference on RF Superconductivity},
  number       = {21},
  publisher    = {JACoW Publishing, Geneva, Switzerland},
  month        = {09},
  year         = {2023},
  issn         = {2673-5504},
  isbn         = {978-3-95450-234-9},
  doi          = {10.18429/JACoW-SRF2023-WEPWB054},
  url          = {https://jacow.org/srf2023/papers/wepwb054.pdf},
  abstract     = {{Jefferson Lab has an ongoing R&D program in plasma processing which just completed a round of production processing in the CEBAF accelerator. Plasma processing is a common technique for removing hydrocarbons from surfaces, which increases the work function and reduces the secondary emission coefficient. Unlike helium processing which relies on ion bombardment of the field emitters, plasma processing uses free oxygen produced in the plasma to break down the hydrocarbons on the surface of the cavity. The initial focus of the effort was processing C100 cavities by injecting RF power into the HOM coupler ports. Results from processing cryomodules in the CEBAF accelerator as well as vertical test results will be presented. The goal will be to improve the operational gradients and the energy margin of the linacs. This work will describe the systems and methods used at JLAB for processing cavities using an argon-oxygen gas mixture as well as a helium-oxygen gas mixture. Before and after plasma processing results will also be presented.}},
}

@inproceedings{giaccone:srf2021-wepcav001,
  author       = {B. Giaccone and M. Martinello and J. Zasadzinski},
  title        = {{Study of the Niobium Oxide Structure and Microscopic Effect of Plasma Processing on the Niobium Surface}},
  booktitle    = {Proc. SRF'21},
% booktitle    = {Proc. 20th International Conference on RF Superconductivity (SRF'21)},
  pages        = {585--589},
  eid          = {WEPCAV001},
  language     = {english},
  keywords     = {plasma, niobium, cavity, background, ECR},
  venue        = {East Lansing, MI, USA},
  series       = {International Conference on RF Superconductivity},
  number       = {20},
  publisher    = {JACoW Publishing, Geneva, Switzerland},
  month        = {10},
  year         = {2022},
  issn         = {2673-5504},
  isbn         = {978-3-95450-233-2},
  doi          = {10.18429/JACoW-SRF2021-WEPCAV001},
  url          = {https://jacow.org/srf2021/papers/wepcav001.pdf},
  abstract     = {{A study of the niobium oxide structure is presented here, with particular focus on the niobium suboxides. Multiple steps of argon sputtering and XPS measurements were carried out until the metal surface was exposed. The sample was then exposed to air and the oxide regrowth was studied. In addition, three Nb samples prepared with different surface treatments were studied before and after being subjected to plasma processing. The scope is investigating the microscopic effect that the reactive oxygen contained in the glow discharge may have on the niobium surface. This study suggests that the Nb₂O₅ thickness may increase, although no negative change in the cavity performance is measured since the pentoxide is a dielectric.}},
}

@article{ZHANG2019143,
title = {The mechanism study of mixed Ar/O2 plasma-cleaning treatment on niobium surface for work function improvement},
journal = {Applied Surface Science},
volume = {475},
pages = {143-150},
year = {2019},
issn = {0169-4332},
doi = {https://doi.org/10.1016/j.apsusc.2018.12.156},
url = {https://www.sciencedirect.com/science/article/pii/S0169433218334949},
author = {Zhiyan Zhang and Zongbiao Ye and Zhijun Wang and Fujun Gou and Bizhou Shen and Andong Wu and Yuan He and Pingni He and Hongbin Wang and Bo Chen and Jianjun Chen and Kun Zhang and Jianjun Wei}
}

@misc{Biagi_LXCat,
  %author       = {Biagi database},
  title        = {Biagi database, electron impact cross section},
  howpublished = {\url{https://lxcat.net}},
  note         = {retrieved on June 25, 2025},
}

@article{OLRY2006197,
title = {Development of a beta 0.12, 88MHz, quarter-wave resonator and its cryomodule for the SPIRAL2 project},
journal = {Physica C: Superconductivity},
volume = {441},
number = {1},
pages = {197-200},
year = {2006},
note = {Proceedings of the 12th International Workshop on RF Superconductivity},
issn = {0921-4534},
doi = {https://doi.org/10.1016/j.physc.2006.03.030},
url = {https://www.sciencedirect.com/science/article/pii/S092145340600178X},
author = {G. Olry and J.-L. Biarrotte and S. Blivet and S. Bousson and C. Commeaux and C. Joly and T. Junquera and J. Lesrel and E. Roy and H. Saugnac and P. Szott}
}

@article{10.1063/1.1769607,
    author = {Hagelaar, G. J. M. and Hassouni, K. and Gicquel, A.},
    title = {Interaction between the electromagnetic fields and the plasma in a microwave plasma reactor},
    journal = {Journal of Applied Physics},
    volume = {96},
    number = {4},
    pages = {1819-1828},
    year = {2004},
    month = {08},
    issn = {0021-8979},
    doi = {10.1063/1.1769607},
    url = {https://doi.org/10.1063/1.1769607},
    eprint = {https://pubs.aip.org/aip/jap/article-pdf/96/4/1819/18712225/1819_1_online.pdf},
}

@article{Hassouni_2010,
doi = {10.1088/0022-3727/43/15/153001},
url = {https://doi.org/10.1088/0022-3727/43/15/153001},
year = {2010},
month = {mar},
publisher = {},
volume = {43},
number = {15},
pages = {153001},
author = {Hassouni, K and Silva, F and Gicquel, A},
title = {Modelling of diamond deposition microwave cavity generated plasmas},
journal = {Journal of Physics D: Applied Physics}
}

@article{GOMEZMARTINEZ201737,
title = {Final results of power conditioning of SPIRAL 2 couplers},
journal = {Nuclear Instruments and Methods in Physics Research Section A: Accelerators, Spectrometers, Detectors and Associated Equipment},
volume = {870},
pages = {37-42},
year = {2017},
issn = {0168-9002},
doi = {https://doi.org/10.1016/j.nima.2017.07.004},
url = {https://www.sciencedirect.com/science/article/pii/S0168900217307155},
author = {Y. Gómez Martínez and M. Baylac and P. Boge and T. Cabanel and P. De Lamberterie and J. Giraud and F. Vezzu and F. Chatelet and C. Joly and J. Lesrel and D. Longuevergne and R. Martret and G. Olry and L. Renard and P. Bosland and C. Marchand and L. Maurice and O. Piquet and P.-E. Bernaudin and R. Ferdinand}
}

@Article{app15137361,
AUTHOR = {Sammut, Stephen},
TITLE = {A Comprehensive Review of Plasma Cleaning Processes Used in Semiconductor Packaging},
JOURNAL = {Applied Sciences},
VOLUME = {15},
YEAR = {2025},
NUMBER = {13},
ARTICLE-NUMBER = {7361},
URL = {https://www.mdpi.com/2076-3417/15/13/7361},
ISSN = {2076-3417},
DOI = {10.3390/app15137361}
}

@article{10.1063/1.1597367,
    author = {Antoine, C. Z. and Berry, S.},
    title = {H in Niobium: Origin And Method Of Detection},
    journal = {AIP Conference Proceedings},
    volume = {671},
    number = {1},
    pages = {176-189},
    year = {2003},
    month = {07},
    issn = {0094-243X},
    doi = {10.1063/1.1597367},
    url = {https://doi.org/10.1063/1.1597367},
    eprint = {https://pubs.aip.org/aip/acp/article-pdf/671/1/176/12002955/176_1_online.pdf},
}

@article{10.1116/1.569186,
author = {Wertheimer, M.R. and Bailon, J.‐P.},
title = {Reduction of niobium pentoxide in a hydrogen discharge},
journal = {Journal of Vacuum Science and Technology},
volume = {14},
number = {2},
pages = {699-704},
year = {1977},
month = {03},
issn = {0022-5355},
doi = {10.1116/1.569186},
url = {https://doi.org/10.1116/1.569186},
eprint = {https://pubs.aip.org/avs/jvst/article-pdf/14/2/699/12152270/699_1_online.pdf},
}

@inproceedings{chen2003langmuir,
  title={Langmuir probe diagnostics},
  author={Chen, Francis F},
  booktitle={IEEE-ICOPS Meeting, Jeju, Korea},
  volume={2},
  number={6},
  year={2003},
  organization={Citeseer}
}

@article{10.1063/1.1149862,
    author = {Hoegy, Walter R. and Brace, Larry H.},
    title = {Use of Langmuir probes in non-Maxwellian space plasmas},
    journal = {Review of Scientific Instruments},
    volume = {70},
    number = {7},
    pages = {3015-3024},
    year = {1999},
    month = {07},
    issn = {0034-6748},
    doi = {10.1063/1.1149862},
    url = {https://doi.org/10.1063/1.1149862},
    eprint = {https://pubs.aip.org/aip/rsi/article-pdf/70/7/3015/19326362/3015_1_online.pdf},
}

@inproceedings{cheneyLCWS,
    author  = {C. Cheney and al},
    title   = "Plasma processing development for SPIRAL2 quarter-wave resonators: experimental and simulation studies",
    publisher = "presented at \emph{2024 International Workshop on Future Linear Colliders (LCWS2024)}, Tokyo, 10 Jul. 2024",
    url = {https://agenda.linearcollider.org/event/10134/contributions/54694/}
}

@misc{inficon,
	author = {inficon.com},
	title = {STM-2 USB Rate Thickness Monitor Operating Manual},
	howpublished = {\url{https://www.inficon.com/media/4289/download/STM-2-USB-Rate-Thickness-Monitor-Operating-Manual-(English).pdf?v=2&inline=true&language=en}},
	year = {},
	note = {[Accessed 03-03-2026]},
}


\end{document}